\documentclass[iop]{emulateapj}

\usepackage{natbib}
\usepackage{aas_macros}
\usepackage[dvips]{color}

\def\e{\mathrm{e}}

\newcommand{\bracket}[1]{\left<#1\right>}

\slugcomment{Draft version \today}

\shorttitle{Impacts of collective neutrino oscillations on supernova
  explosions}
\shortauthors{SUWA ET AL.}

\begin{document}

\title{Impacts of Collective Neutrino Oscillations on Core-Collapse
  Supernova Explosions}

\author{
Yudai Suwa\altaffilmark{1},
Kei Kotake\altaffilmark{2,3}, 
Tomoya Takiwaki\altaffilmark{3},
Matthias Liebend\"orfer\altaffilmark{4},
and Katsuhiko Sato\altaffilmark{5,6}}

\altaffiltext{1}{Yukawa Institute for Theoretical Physics, Kyoto
  University, Oiwake-cho, Kitashirakawa, Sakyo-ku, Kyoto, 606-8502,
  Japan}
\altaffiltext{2}{Division of Theoretical Astronomy, National
  Astronomical Observatory of Japan, Mitaka, Tokyo 181-8588, Japan}
\altaffiltext{3}{Center for Computational Astrophysics, National
  Astronomical Observatory of Japan, Mitaka, Tokyo 181-8588, Japan}
\altaffiltext{4}{Department of Physics, University of Basel,
  Klingelbergstr. 82, CH-4056 Basel, Switzerland} 
\altaffiltext{5}{The Institute for the Physics and Mathematics of the
  Universe, the University of Tokyo, Kashiwa, Chiba, 277-8568, Japan}
\altaffiltext{6}{National Institutes of Natural Sciences, Kamiyacho Central Place 2F, 
4-3-13 Toranomon, Minato-ku, Tokyo, 105-0001, Japan}

\email{suwa@yukawa.kyoto-u.ac.jp}

\begin{abstract}
By performing a series of one- and two-dimensional (1-, 2D)
hydrodynamic simulations with spectral neutrino transport, we study
possible impacts of collective neutrino oscillations on the dynamics
of core-collapse supernovae.  To model the spectral swapping which is
one of the possible outcome of the collective neutrino oscillations,
we parametrize the onset time when the spectral swap begins, the
radius where the spectral swap occurs, and the threshold energy above
which the spectral interchange between heavy-lepton neutrinos and
electron/anti-electron neutrinos takes place, respectively. By doing
so, we systematically study how the neutrino heating enhanced by the
spectral swapping could affect the shock evolution as well as the
matter ejection.  We also investigate the progenitor dependence using
a suite of progenitor models (13, 15, 20, and 25 $M_\odot$).  We find
that there is a critical heating rate induced by the spectral swapping
to trigger explosions, which significantly differs between the
progenitors.  The critical heating rate is generally smaller for 2D
than 1D due to the multidimensionality that enhances the neutrino
heating efficiency. For the progenitors employed in this paper, the
final remnant masses are estimated to range in 1.1-1.5$M_\odot$. For
our 2D model of the $15M_\odot$ progenitor, we find a set of the
oscillation parameters that could account for strong supernova
explosions ($\sim 10^{51}$ erg), simultaneously leaving behind the
remnant mass close to $\sim 1.4 M_\odot$.
\end{abstract}


\section{Introduction}
\label{sec:intro}
Although the explosion mechanism of core-collapse supernovae is not
completely understood yet, current multi-dimensional (multi-D)
simulations based on refined numerical models show several promising
scenarios. Among the candidates are the neutrino heating mechanism
aided by convection and standing accretion shock instability (SASI)
\cite[e.g.,][]{mare09,brue09,suwa10}, the acoustic mechanism
\citep{burr07a}, or the magnetohydrodynamic (MHD) mechanism
\cite[e.g.,][]{kota04a,kotarev,ober06,burr07b,taki09}. Probably the
best-studied one is the neutrino heating mechanism, whose basic
concept was first proposed by \citet{colg66}, and later reinforced by
\citet{beth85} to take a currently prevailing delayed form.

An important lesson from the multi-D simulations mentioned above is
that hydrodynamic motions associated with convective overturn
\citep{hera94,burr95,jank96,frye02,frye04b} as well as the SASI
\citep[e.g.,][]{blon03,sche06,ohni06,fogl07,murp08,iwak08,guil10,fern10}
can help the onset of the neutrino-driven explosion, which otherwise
fails generally in spherically symmetric (1D) simulations
\citep{lieb01,ramp02,thom03,sumi05}.  This is mainly because the
accretion timescale of matter in the gain region can be longer than in
the 1D case, which enhances the strength of neutrino-matter coupling
there.

In fact, the neutrino-driven explosions have been obtained in the
following state-of-the-art two-dimensional (2D) simulations.  Using
the MuDBaTH code which includes one of the best available neutrino
transfer approximations, \cite{bura06} firstly reported explosions for
a non-rotating low-mass ($11.2 M_\odot$) progenitor of \cite{woos02},
and then for a $15 M_{\odot}$ progenitor of \citet{woos95} with a
moderately rapid rotation imposed \citep{mare09}.  By implementing a
multi-group flux-limited diffusion algorithm to the CHIMERA code
\cite[e.g.,][]{brue09}, \citet{yaku10} obtained explosions for a
non-rotating $12 M_{\odot}$ and 25$M_{\odot}$ progenitor of
\citet{woos02}.  More recently, \cite{suwa10} pointed out that a
stronger explosion is obtained for a rapidly rotating $13M_\odot$
progenitor of \cite{nomo88} compared to the corresponding non-rotating
model, in which the isotropic diffusion source approximation (IDSA)
for the spectral neutrino transport \citep{lieb09} is implemented in
the ZEUS code.

However, this success opens further new questions. First of all, the
explosion energies obtained in these simulations are typically
underpowered by one or two orders of magnitudes to explain the
canonical supernova kinetic energy ($\sim 10^{51}$ erg).  Moreover,
the softer nuclear equation of state (EOS), such as of the
\citet{latt91} (LS) EOS with an incompressibility $K=180$ MeV at
nuclear densities is employed in those simulations.  On top of
evidence that favors a stiffer EOS based on nuclear experimental data
\citep{shlo06}, the soft EOS may not account for the recently observed
massive neutron star of $\sim 2 M_{\odot}$ \citep{demo10} \citep[see
  the maximum mass for the LS180 EOS in][]{ocon11}. With a stiffer
EOS, the explosion energy may be even lower as inferred from
\citet{mare09} who did not obtain the neutrino-driven explosion for
their model with $K=263$ MeV.  What is then missing furthermore?  We
may get the answer by going to 3D simulations \citep{nord10} or by
taking into account new ingredients, such as exotic physics in the
core of the protoneutron star \citep{sage09}, viscous heating by the
magnetorotational instability \citep{thom05,masa11}, or energy
dissipation via Alfv\'en waves \citep{suzu08}.

Joining in these efforts, we explore in this study the possible
impacts of collective neutrino oscillations on energizing the
neutrino-driven explosions. The collective neutrino oscillations,
i.e. neutrinos of all energies that oscillate almost in phase, are
attracting great attention, because they can induce dramatic
observable effects such as a spectral split or swap \cite[e.g.,][and
  see references therein]{raff07,duan08,dasg08}. They are predicted to
emerge as a distinct feature in their energy spectra \cite[see][for
  reviews of the rapidly growing research field and collective
  references therein]{duan10,dasg10}. Among a number of important
effects possibly created by the self-interaction, we choose to
consider the effect of spectral splits between electron- ($\nu_e$),
anti-electron neutrinos ($\bar{\nu}_e$), and heavy lepton neutrinos
($\nu_x$, i.e., $\nu_{\mu}$, $\nu_{\tau}$ and their anti-particles)
above a threshold energy (e.g., \citet{fogli07}).  Since $\nu_x$'s
have higher average energies than the other species in the postbounce
phase, the neutrino flavor mixing would increase the effective
energies of $\nu_e$ and $\bar{\nu}_e$, and hence increase the neutrino
heating rates in the gain region.  A formalism to treat the neutrino
oscillation in the Boltzmann neutrino transport is given in
\citet{yama00,stra05}, but difficult to implement.  To just mimic the
effects in this study, we perform the spectral swap by hand as a first
step.  By changing the average neutrino energy,
$\bracket{\epsilon_{\nu_{x}}}$, as well as the position of the
neutrino spheres ($R_{\nu_x}$) in a parametric manner, we hope to
constrain the parameter regions spanned by
$\bracket{\epsilon_{\nu_{x}}}$ and $R_{\nu_x}$ in which the additional
heating given by the collective neutrino oscillations could have
impacts on the explosion dynamics. Our strategy is as follows.  By
performing a number of 1D simulations, we will firstly constrain the
parameter regions to some extent.  Here we also investigate the
progenitor dependence using a suite of progenitor models (13, 15, 20,
and 25 $M_\odot$).  After squeezing the condition in the 1D
computations, we include the flavor conversions in 2D simulations to
see their impacts on the dynamics, and we also discuss how the
critical condition for the collective effects in 1D can be subject to
change in 2D.
 
The paper opens with descriptions of the initial models and the
numerical methods focusing how to model the collective neutrino
oscillations (Section 2).  The main results are shown in Section 3.
We summarize our results and discuss their implications in Section 4.

\section{Numerical Methods}

\subsection{Hydrodynamics}
The employed numerical methods are essentially the same as those in
our previous paper \citep{suwa10}.  For later convenience, we briefly
summarize them in the following.  The basic evolution equations are
written as,
\begin{equation}
  \frac{\mathrm{d}\rho}{\mathrm{d}t}+\rho\nabla\cdot\mathbf{v}=0,
\end{equation}
\begin{equation}
  \rho \frac{\mathrm{d}\mathbf{v}}{\mathrm{d}t}=-\nabla P -\rho \nabla \Phi,
\end{equation}
\begin{equation}
\frac{\mathrm{d}e^*}{\mathrm{d}t}+
\nabla \cdot
\left[\left(e^* + P\right) \mathbf{v}\right]= -\rho \mathbf{v} \cdot \nabla  \Phi+ Q_{E}
\label{eq:e},
\end{equation}
\begin{equation}
\frac{{\mathrm{d}Y_e}}{\mathrm{d}t} = Q_N,
\label{eq:ye}
\end{equation}
\begin{equation}
  \bigtriangleup{\Phi} = 4\pi G \rho,
\end{equation}
where $\rho, \mathbf{v}, P, \mathbf{v}, e^*, \Phi$, are density, fluid
velocity, gas pressure including the radiation pressure of neutrinos,
total energy density, gravitational potential, respectively. The time
derivatives are Lagrangian. As for the hydro solver, we employ the
ZEUS-2D code \citep{ston92} which has been modified for core-collapse
simulations \citep[e.g.,][]{suwa07a,suwa07b,suwa09a,taki09}. $Q_{E}$
and $Q_{N}$ (in Equations (\ref{eq:e}) and (\ref{eq:ye})) represent
the change of energy and electron fraction ($Y_e$) due to the
interactions with neutrinos. To estimate these quantities, we
implement spectral neutrino transport using the isotropic diffusion
source approximation (IDSA) scheme \citep{lieb09}. The IDSA scheme
splits the neutrino distribution into two components, both of which
are solved using separate numerical techniques. We apply the so-called
ray-by-ray approach in which the neutrino transport is solved along a
given radial direction assuming that the hydrodynamic medium for the
direction is spherically symmetric. Although the current IDSA scheme
does not yet include $\nu_x$ and the inelastic neutrino scattering
with electrons, these simplifications save a significant amount of
computational time compared to the canonical Boltzmann solvers (see
\cite{lieb09} for more details). Following the prescription in
\cite{muel10}, we improve the accuracy of the total energy
conservation by using a conservation form in equation (\ref{eq:e}),
instead of solving the evolution of internal energy as originally
designed in the ZEUS code. Numerical tests are presented in Appendix
A.

The simulations are performed on a grid of 300 logarithmically spaced
radial zones from the center up to 5000 km and 128 equidistant angular
zones covering $0\le\theta\le\pi$ for two-dimensional simulations. For
the spectral transport, we use 20 logarithmically spaced energy bins
reaching from 3 to 300 MeV.

\subsection{Spectral swapping}

As mentioned in \S\ref{sec:intro}, we introduce a spectral interchange
from heavy-lepton neutrinos ($\nu_\mu$, $\nu_\tau$ and their
antineutrinos, collectively referred as $\nu_x$ hereafter) to
electron-type neutrinos and antineutrinos, namely $\nu_x \rightarrow
\nu_e$ and $\bar{\nu}_x \rightarrow \bar{\nu}_e$. Instead of solving
the transport equations for $\nu_x$, we employ the so-called
light-bulb approximation and focus on the optically thin region
outside the neutrinosophere \citep[e.g.,][]{jank96,ohni06}.

According to \cite{duan10}, we set the threshold energy,
$\epsilon_{th}$, to be 9 MeV, above which the spectral swap takes
place. Below the threshold, the neutrino heating is estimated by the
spectral transport via the IDSA scheme. Above the threshold, the
heating rate is replaced by
\begin{equation}
Q_E\propto \int^\infty_{\epsilon_{th}}
d\epsilon_\nu~\epsilon^3\left[j(\epsilon_\nu)+\chi(\epsilon_\nu)\right]
f_{\nu}(r,\epsilon_\nu),
\end{equation}
where $j$ and $\chi$ are the neutrino emissivity and absorptivity,
respectively, and $f_{\nu}(r,\epsilon_\nu)$ corresponds to the
neutrino distribution function for $\nu_x$ with $\epsilon_\nu$ being
energies of electron neutrinos and antineutrinos. In the light-bulb
approach, it is often approximated by the Fermi-Dirac distribution
with a vanishing chemical potential \citep[e.g.,][]{ohni06} as,
\begin{equation}
f_{\nu}(r,\epsilon_\nu)=\frac{1}{e^{\epsilon_\nu/kT_{\nu_x}}+1}g(r),
\label{eq:f}
\end{equation}
where $k$, $T_{\nu_x}$ are the Boltzmann constatn and the neutrino
temperature, respectively. $g(r)$ is the geometric factor,
$g(r)=1-\left[1-(R_{\nu_x}/r)^2\right]^{1/2}$ which is taken into
account for the normalization, with $R_{\nu_x}$ being the radius of
the neutrinosphere. The neutrino luminosity of $\nu_x$ at the infinity
is the given as
\begin{equation}
L_{\nu_x}=2.62\times10^{52}\left(\frac{\bracket{\epsilon_{\nu_x}}}{15~\mathrm{MeV}}\right)^4\left(\frac{R_{\nu_x}}{30~\mathrm{km}}\right)^2~\mathrm{erg~s^{-1}},
\label{eq:Lnu}
\end{equation}
where $\bracket{\epsilon_{\nu_x}}=\int^\infty_0
d\epsilon_{\nu_x}\epsilon_{\nu_x}^3
f_\nu(\epsilon_{\nu_x})/\int^\infty_0
d\epsilon_{\nu_x}\epsilon_{\nu_x}^2 f_\nu(\epsilon_{\nu_x})$ is the
average energy of emitted neutrinos.  The position where the spectral
swapping sets in is fixed at 100 km (around the gain radius) and the
onset time is varied as a parameter, $t_s=$100, 200, and 300 ms after
bounce.

In fact, the threshold energy depends on the neutrino luminosities,
spectra and oscillation parameters \cite[see, e.g.,][and references
  therein]{duan10} with conserved net $\nu_e$ flux (i.e., the lepton
number conservation). However, the conservation of lepton number is
too complicated to satisfy in the dynamical simulation because the
neutrino spectrum and the luminosity evolve with time. In order to
focus on the hydrodynamic features affected by the spectral modulation
induced by the swapping, we simplify just a single threshold energy in
this work.

To summarize, the parameters that we use to mimic the spectral
swapping are the following three items, (i) $R_{\nu_x}$ which is the
radius of the neutrinosphere of $\nu_x$, (ii)
$\bracket{\epsilon_{\nu_x}}$ which is the average energy of $\nu_x$,
and (iii) $t_s$ which is the time when the spectral swapping sets in.

\section{Result}

\subsection{One-dimensional models}

\subsubsection{1D without spectral swapping}

In this subsection, we first outline the 1D collapse dynamics without
spectral swapping. We take a 13 $M_{\odot}$ progenitor \citep{nomo88}
as a reference.

\begin{figure}[tbp]
\includegraphics[width=0.5\textwidth]{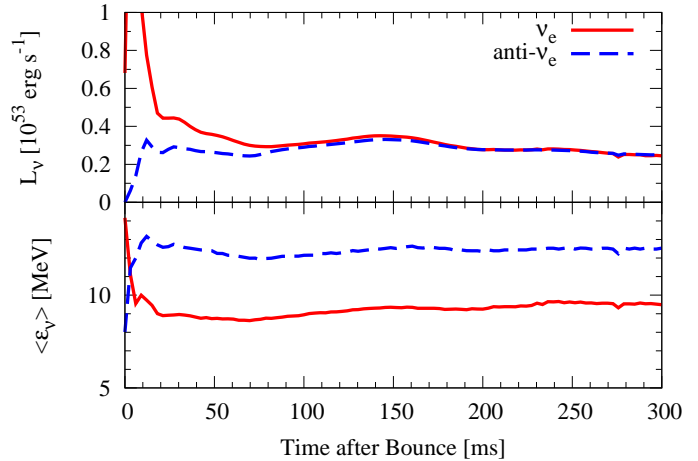}
\caption{Time evolution of the neutrino luminosity (top panel) and
  average energy (bottom panel) for $\nu_e$ (red-solid line) and
  $\bar\nu_e$ (blue-dashed line).}
\label{fig:neu_lum}
\end{figure}

\begin{figure}[tbp]
\includegraphics[width=0.5\textwidth]{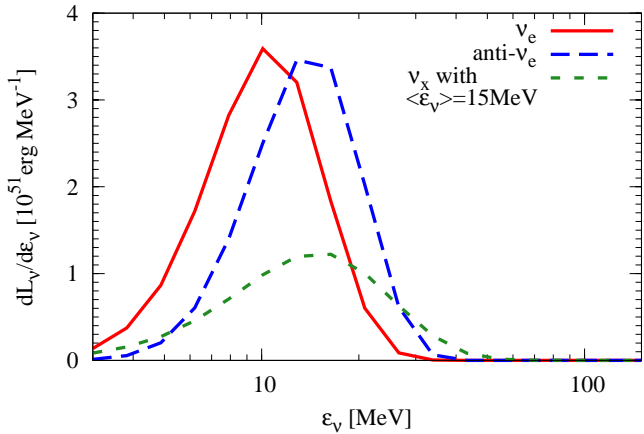}
\caption{The neutrino luminosity spectrum of $\nu_e$ (red-solid line)
  and $\bar\nu_e$ (blue-dashed line) without the spectral swapping at
  100 ms after the bounce. For comparison, we show the injected
  luminosity spectrum of $\nu_x$ with
  $\bracket{\epsilon_{\nu_x}}=15$MeV, which will be swapped with the
  original spectrum of $\nu_e$ and $\bar\nu_e$ at $\epsilon_\nu>$ 9
  MeV for models including spectral swapping.}
\label{fig:spec}
\end{figure}

At around 112 ms after the onset of gravitational collapse, the bounce
shock forms at a radius of $\sim$ 10 km with an enclosed mass of $\sim
0.7M_\odot$\footnote{Note that $0.7M_\odot$ is rather high value that
  is due to approximations employed in our simulation. We omit the
  electron scattering by neutrinos and general-relativistic effects,
  which lead smaller inner core mass at bounce
  \citep[see][]{lieb01,thom03}. In addition, more improved electron
  capture treatment would lead even smaller \citep{lang03}.}.  The
central density at this time is $\rho_c=3.6\times 10^{14}$ g
cm$^{-3}$. The shock propagates outwards but finally stalls at a
radius of $\sim$ 100 km.  Due to the decreasing accretion rate through
the stalled shock, the shock can be still pushed outward. However,
after some time, the shock radius begins to shrink.  The ratio of the
advection timescale, $\tau_\mathrm{adv}$, and the heating timescale,
$\tau_\mathrm{heat}$, is an important indicator for the criteria of
neutrino driven explosion \citep{bura06,mare09,suwa10}. In our 1D
simulations, $\tau_\mathrm{adv}/\tau_\mathrm{heat}$ is generally
smaller than unity in the postbounce phase. This is the reason why our
1D simulations do not yield a delayed explosion.  This also the case
for the other progenitors (15, 20, and 25 $M_\odot$) investigated in
this study.  As for the accretion phase (later than $\sim$ 50 ms after
the bounce), the typical neutrino luminosity at $r=5000$ km is
$3\times 10^{52}$ erg s$^{-1}$ for both $\nu_e$ and $\bar\nu_e$, and
the typical average energy is $\bracket{\epsilon_{\nu_e}}\approx 9$
MeV and $\bracket{\epsilon_{\bar\nu_e}}\approx 12$ MeV as shown in
Figure \ref{fig:neu_lum}. Figure \ref{fig:spec} indicates the
resultant neutrino luminosity spectrum at 100 ms after the bounce.

\subsubsection{1D with spectral swapping}\label{sec:1D}

The investigated models with the spectral swapping are summarized in
Table \ref{tab:models}. As already mentioned, the model parameters are
the neutrinosphere radius ($R_{\nu_x}$), the average energy of
neutrinos ($\bracket{\epsilon_{\nu_x}}$), and the onset time of the
spectral swapping ($t_s$). The model names include these parameters;
``NH13'' represents the progenitor model, ``R..'' represents
$R_{\nu_x}$ in units of km, ``E..'' represents
$\bracket{\epsilon_{\nu_x}}$ in MeV, ``T..''  represents $t_s$ in ms,
and the last letter ``S'' represents 1D (spherical symmetry).

\begin{figure}[tbp]
\includegraphics[width=0.5\textwidth]{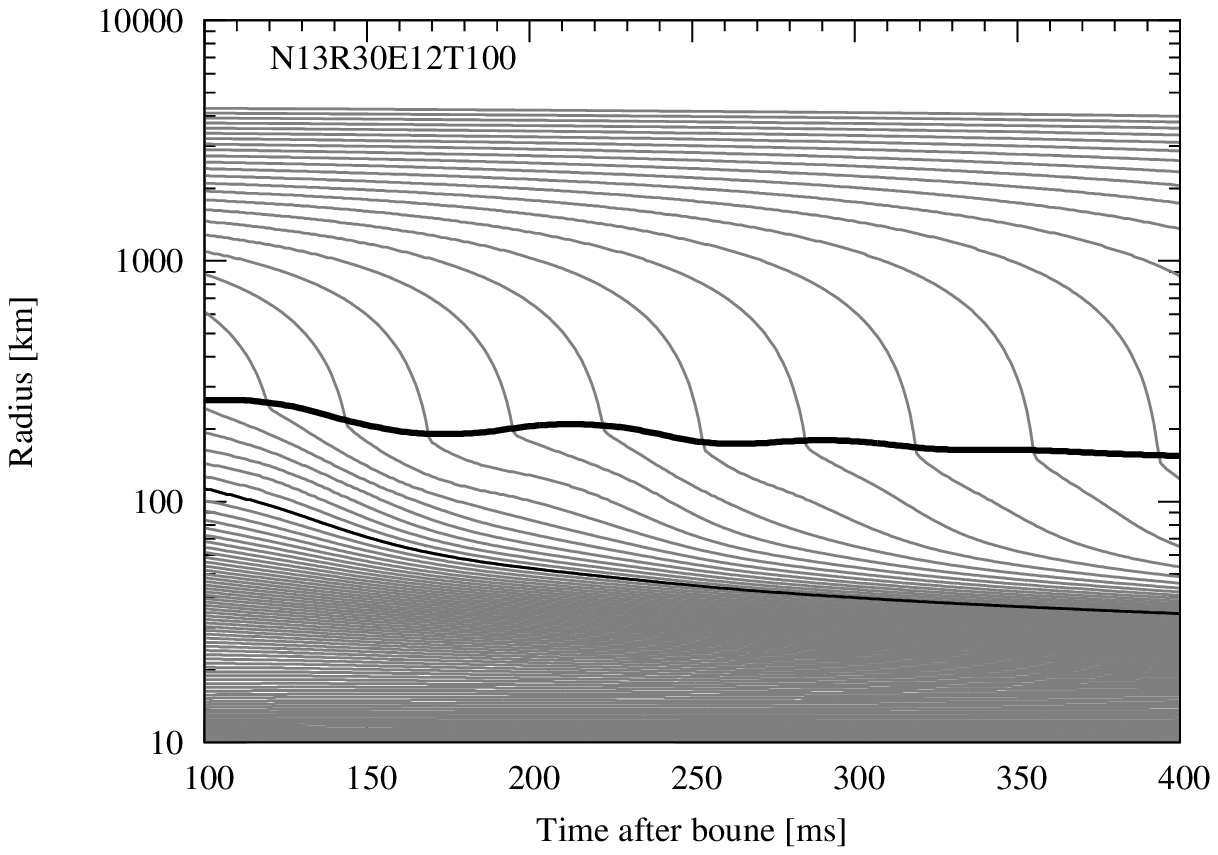}
\includegraphics[width=0.5\textwidth]{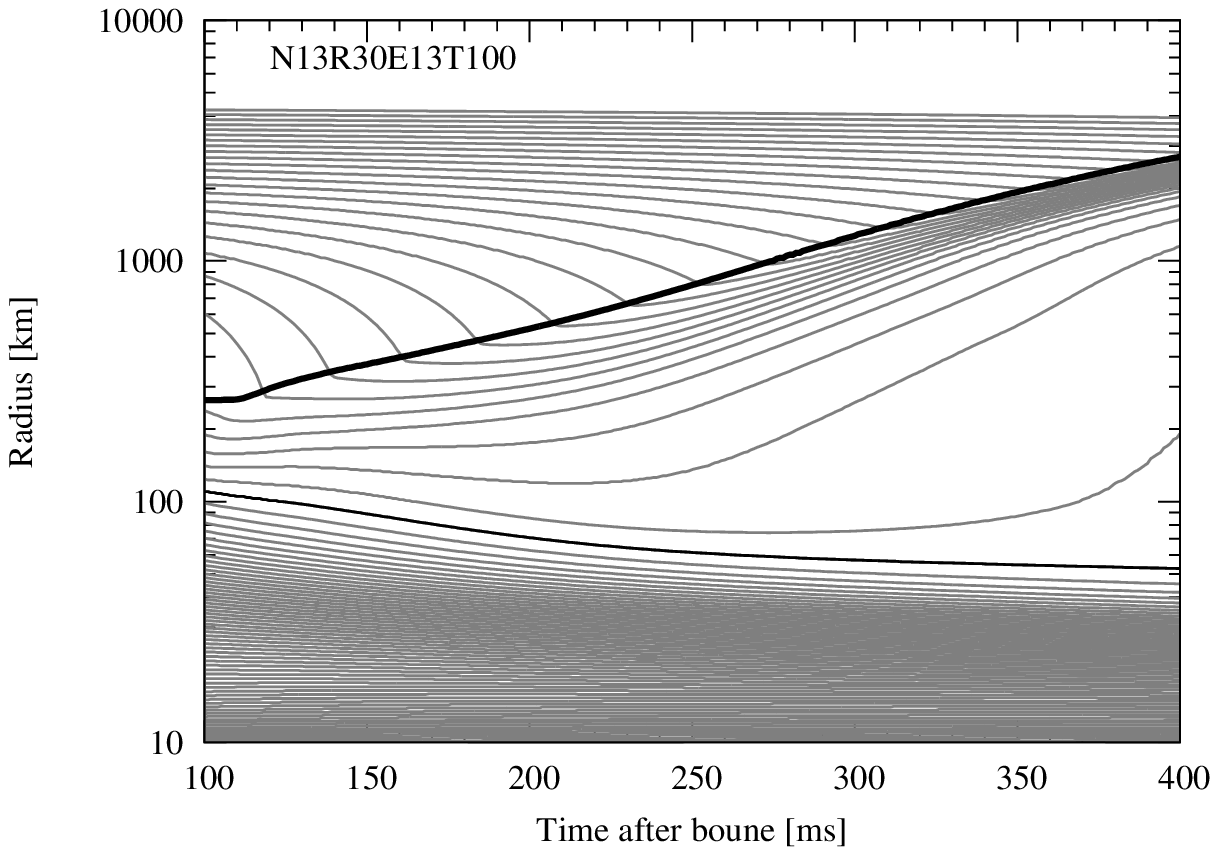}
\caption{Time evolution of mass shells for NH13R30E12T100S (top) and
  NH13R30E13T100S (bottom). The Black thin line corresponds to $1.2
  M_\odot$ and the black thick line represents the shock wave
  position, respectively.  The difference between these panels is the
  average energies of neutrinos, $\bracket{\epsilon_{\nu_x}}=12$ MeV
  for the top panel and 13 MeV for the bottom panel.}
\label{fig:mass_shell}
\end{figure}

Figure \ref{fig:mass_shell} presents the time evolution of the mass
shells for models NH13R30E12T100S and NH13R30E13T100S. The difference
between these panels is the average energies of neutrinos,
$\bracket{\epsilon_{\nu_x}}=12$ MeV for the top panel and 13 MeV for
the bottom panel. The thick solid lines represent the radial position
of shock waves.  Regardless of a small difference of
$\bracket{\epsilon_{\nu_x}}$, model NH13R30E13T100S shows a shock
expansion after the manual spectral swapping is switched on (see the
thick line in the bottom panel of Figure \ref{fig:mass_shell}), while
the stalled shock does not revive for model NH13R30E12T100S (top
panel).  This suggests that there is a critical condition for the
successful explosion induced by the spectral swapping.  In the bottom
panel, the regions enclosing the mass of $M_r \sim 1.2 M_\odot$ (thin
black line) corresponds to the so-called mass cut, which could be
interpreted as the final mass of the remnant.  The fact that a clear
mass cut emerges in model NH13R30E13T100S indicates that a neutron
star will be left behind in this model.  Such a definite mass-cut has
been observed in \citet{kita06} who reported a successful
neutrino-driven explosion (in 1D) for a lighter progenitor star, which
is, however, difficult to realize for more massive stars in 2D (e.g.,
Figure 2 in \citet{mare09} and Figure 1 in \citet{suwa10}).

\begin{figure}[tbp]
\includegraphics[width=0.5\textwidth]{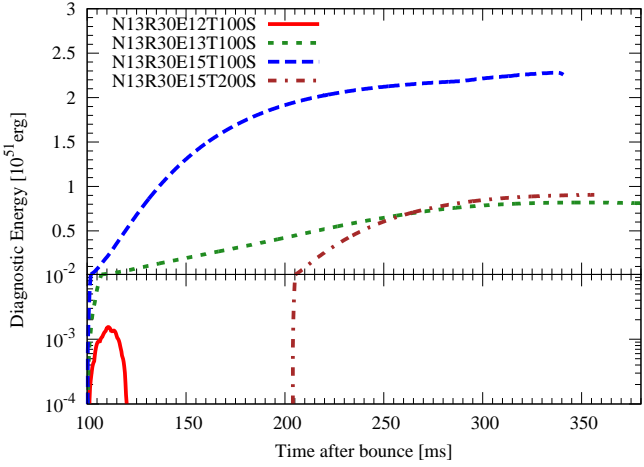}
\caption{Diagnostic energies as functions of time. Red-solid,
  green-dotted, blue-dashed and brown-dot-dashed lines correspond to
  models NH13R30E12T100S, NH13R30E13T100S, NH13R30E15T100S, and
  NH13R30E15T200S, respectively. The $\nu_x$ average energies of three
  lines other than brown-dot-dashed line differ from each other,
  i.e. $\bracket{\epsilon_{\nu_x}}=$ 12 MeV (red solid), 13 MeV (green
  dotted), and 15 MeV (blue dashed), respectively. As for the
  brown-dot-dashed line, only the onset time of spectral swapping is
  different from the blue-dashed line.  The red-solid line shows only
  an oscillation, while the other lines show increasing diagnostic
  energy.}
\label{fig:time_ev}
\end{figure}

As a tool to measure the strength of an explosion, we define a {\em
  diagnostic} energy that refers to
\begin{equation}
E_\mathrm{diag}=\int_D dV \left(\frac{1}{2}\rho |v|^2+e-\rho\Phi \right),
\label{eq:Ediag}
\end{equation}
where $e$ is internal energy, $D$ represents the domain in which the
integrand is positive.  Figure \ref{fig:time_ev} shows the time
evolution of $E_\mathrm{diag}$ for some selected models. The
diagnostic energy increases with time for the green-dotted line, which
turns to decrease for the red line, noting that the difference between
the pair of models is $\Delta \bracket{\epsilon_{\nu_x}} = 1$ MeV.
The blue-dashed line (model NH13R30E15T100S) has
$\bracket{\epsilon_{\nu_x}}=15$ MeV and reaches larger
$E_\mathrm{diag}$ than the green line (NH13R30E13T100S;
$\bracket{\epsilon_{\nu_x}}=13$ MeV). On the other hand, the later
injection of the spectral swapping leads to smaller $E_\mathrm{diag}$,
i.e. the brown-dot-dashed line ($t_s$=200 ms) shows smaller
$E_\mathrm{diag}$ than the blue-dashed line ($t_s=$100 ms). For models
that experience earlier spectral swapping with higher neutrino energy,
the diagnostic energy becomes higher in an earlier stage, as it is
expected.

Looking at Figure \ref{fig:time_ev} again, $E_\mathrm{diag}$ for the
exploding models seems to show a saturation with time. These curves
can be fitted by the following function,
\begin{equation}
E_\mathrm{diag}(t)=E_\mathrm{diag}^\infty(1-\e^{-at+b}),
\end{equation}
where $E_\mathrm{diag}^\infty$ is a converging value of
$E_\mathrm{diag}$, $a$ and $b$ are the fitting parameters.  As for
NH13R30E13T100S, $E_\mathrm{diag}^\infty=8.5\times 10^{50}$ erg. This
fitting formula allows us to estimate the final diagnostic energy
especially for the strongly exploding models whose diagnostic energy
we cannot estimate in principle because the shock goes beyond the
computational domains ($r<5000$ km) before the saturation.

\begin{figure}[tbp]
\includegraphics[width=0.5\textwidth]{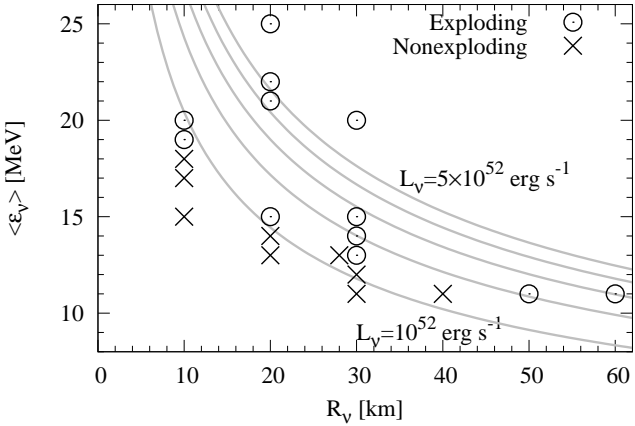}
\caption{Summary of 1D models with $t_s=100$ ms after the
  bounce. Circles indicate the exploding models, while crosses show
  non-exploding models. Gray solid lines correspond to the luminosity
  of $\nu_x$ calculated by Eq. (\ref{eq:Lnu}), which are 1, 2, 3, 4,
  and $5\times 10^{5 2}$ erg s$^{-1}$ from bottom to top.}
\label{fig:param}
\end{figure}

Figure \ref{fig:param} shows the summary of 1D models.  For a given
neutrino luminosity that is determined by $R_{\nu_x}$ and
$\bracket{\epsilon_{\nu_x}}$ (equation (\ref{eq:Lnu})). The gray lines
correspond to the neutrino luminosities determined by the pairs of
$R_{\nu_x}$ and $\bracket{\epsilon_{\nu_x}}$ which is 1 to $5\times
10^{52}$ erg s$^{-1}$ from bottom to top lines. Circles and crosses
correspond to the exploding and non-exploding models, respectively.
Not surprisingly, explosions are more easier to be obtained for higher
neutrino luminosity.

As is well known, the combination of $\bracket{\epsilon_{\nu_x}}$ and
$L_{\nu_x}$ is an important quantity to diagnose the success or
failure of explosions, because the neutrino heating rate in the
so-called gain region, $Q_\nu^+$, is proportional to
$\bracket{\epsilon_{\nu_x}^2} L_{\nu_x}$ (e.g., equation (23) in
\citet{jank01}).

\begin{figure}[tbp]
\includegraphics[width=0.5\textwidth]{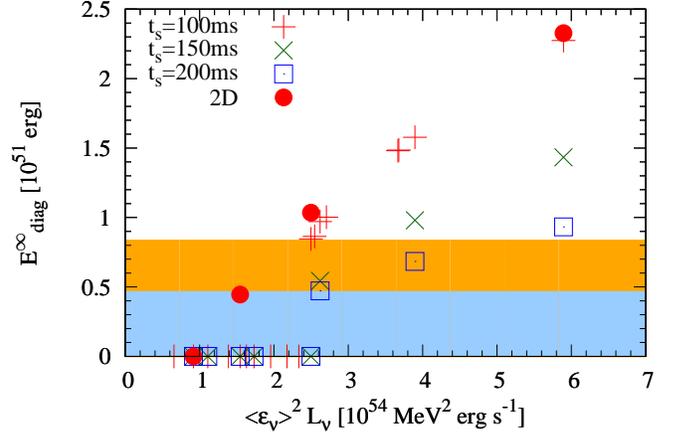}
\caption{Diagnostic energies for exploding models at several hundred
  seconds after the bounce. Red points, green crosses, and blue
  squares correspond to models with $t_s=$100, 150, and 200 ms,
  respectively.  Red circles represent the result of 2D simulation
  (see text for details).}
\label{fig:Ediag}
\end{figure}

Figure \ref{fig:Ediag} shows $E_\mathrm{diag}^\infty$ as a function of
$\bracket{\epsilon_{\nu_x}}^2L_{\nu_x}$. Note in the plot that we set
the horizontal axis not as $\bracket{\epsilon_{\nu_x}^2}L_{\nu_x}$ but
as $\bracket{\epsilon_{\nu_x}}^2L_{\nu_x}$ so that we can deduce the
following dependence more clearly and
easily\footnote{$\bracket{\epsilon_{\nu_x}^2}\left(\equiv\int^\infty_0
  d\epsilon_{\nu_x}\epsilon_{\nu_x}^5f_\nu(\epsilon_{\nu_x})/\int^\infty_0
  d\epsilon_{\nu_x}\epsilon_{\nu_x}^3f_\nu(\epsilon_{\nu_x})\right)$
  and $\bracket{\epsilon_{\nu_x}}$ can be simply connected as
  $\bracket{\epsilon_{\nu_x}^2}=2.1\bracket{\epsilon_{\nu_x}}^2$ for
  the neutrino spectrum of equation (\ref{eq:f}).}. In this figure,
let us first focus on red pluses, green crosses, and blue squares
whose difference is characterized by $t_s$ (2D results (filled
circles) will be mentioned in the later section).  Red ($t_s=$100 ms),
green ($t_s=$150 ms), and blue ($t_s=$200 ms) points have a clear
correlation with $\bracket{\epsilon_{\nu_x}}^2L_{\nu_x}$.  Orange and
light-blue regions represent the non-exploding regions for red and
blue points, respectively.  Both of them show that the minimum
$E_\mathrm{diag}^\infty$ decreases with $t_s$, indicating that the
critical values of $\bracket{\epsilon_\nu}^2L_\nu$ for explosion
sharply depends on $t_s$.  This is because the mass outside the shock
wave gets smaller with time so that the minimum energy to blow up star
gets smaller too.  By the same reason, $E_\mathrm{diag}$ becomes
larger as $t_s$ becomes smaller given the same
$\bracket{\epsilon_{\nu_x}}^2L_{\nu_x}$.  To obtain a larger
$E_\mathrm{diag}^\infty$, the earlier spectral swapping is more
preferential.

\begin{figure}[tbp]
\includegraphics[width=0.5\textwidth]{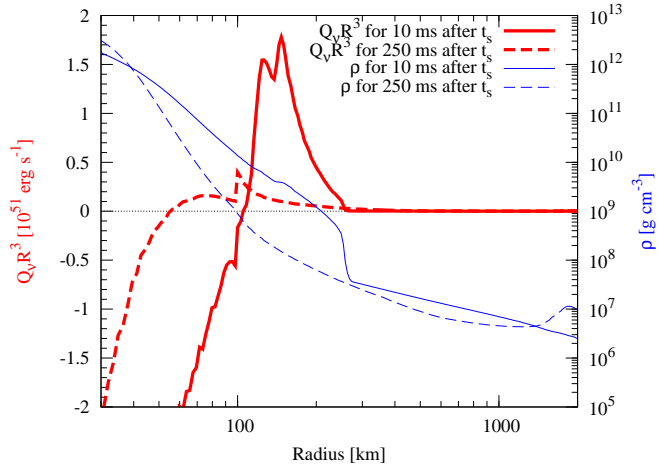}
\caption{Heating rate at 10 ms (red-solid line) and 250 ms (red-dashed
  line) after the bounce for the model NH13R30E13T100S and the density
  profile (blue-solid and dashed lines). As the density decreases due
  to the neutrino driven wind, the heating rate also decreases.}
\label{fig:edot}
\end{figure}

Figure \ref{fig:edot} shows the neutrino heating rate and the density
distribution of NH13R30E13T100S for 10 ms and 250 ms after $t_s$ (=100
ms after the bounce). As the shock wave propagates outward, the
density in the gain region sharply drops (e.g., 100-200km, dashed blue
line), leading to the suppression of the heating rate (dashed red
line).  This is the reason of the saturation in $E_\mathrm{diag}$ as
shown in Figure \ref{fig:time_ev}.

The remnant mass is an important indicator to diagnose the
consequences of the explosion in producing either a neutron star or a
black hole. The last two lines in Table \ref{tab:models} show the
integrated masses in the regions of $\rho\ge 10^{10}$ g cm$^{-3}$ at
$t=t_s$ and $t=\infty$. The latter one is estimated by the fitting as
\begin{equation}
M_{10}(t)=M_{10}^\infty(1+\e^{-ct+d}),
\end{equation}
where $c$ and $d$ are the fitting parameters.  For the exploding
models, $M_{10}^\infty$ becomes generally smaller than
$M_{10}^{t=t_s}$ because of the mass ejection.  Exceptions are weakly
exploding models (NH13R20E15T150S, NH13R20E15T200S, NH13R30E13T100S,
and NH13R50E11T100S), in which the mass accretion continues after
$t_s$ and stops eventually at late time (maximum masses are presented
in Table \ref{tab:models}).  For the nonexploding models, the remnant
mass simply increases with time. Regarding the 13 $M_{\odot}$
progenitor ivestigated in this section, the remnant masses in models
that produce strong explosion ($E_\mathrm{diag}^{\infty}\gtrsim
10^{51}$ erg), are considerably smaller (1.1-1.2 $M_\odot$) if
compared to the typical mass as of observed neutron stars $\sim 1.4
M_\odot$ \citep{latt07}.  This may simply reflect the light iron core
($\sim 1.26 M_{\odot}$) inherent to the progenitor or the existence of
mass accretion induced by the matter fallback after the explosion. Now
we move on to investigate the progenitor dependence in the next
section.

\begin{deluxetable*}{lccccccccc}
\tablecaption{1D simulations}
\tablecolumns{10}
\tablewidth{0pc}
\tabletypesize{\scriptsize}
\tablehead{
\colhead{Model} & \colhead{Dimension} & \colhead{$R_\nu$} & \colhead{$\bracket{\epsilon_{\nu_x}}$} & \colhead{$L_\nu$} & \colhead{$t_s$}  & \colhead{Explosion} & \colhead{$E_\mathrm{diag}^ \infty$} & \colhead{$M_{10}^{t=t_s}$} & \colhead{$M_{10}^\infty$} \\
\colhead{} & \colhead{} & \colhead{[km]} & \colhead{[MeV]}  & \colhead{[10$^{52} $erg s$^{-1}$]} & \colhead{[ms]} & \colhead{} & \colhead{[$10^{51}$ erg]} & \colhead{[$M_\odot$]} & \colhead{[$M_\odot$]}
}
\startdata
NH13R10E15T100S   & 1D & 10 & 15MeV &  0.29 & 100 & No            & ---  & 1.18 & --- \\
NH13R10E17T100S   & 1D & 10 & 17MeV &  0.48 & 100 & No            & ---  & 1.18 & --- \\
NH13R10E18T100S   & 1D & 10 & 18MeV &  0.60 & 100 & No            & ---  & 1.18 & --- \\
NH13R10E19T100S   & 1D & 10 & 19MeV &  0.75 & 100 & {\bf Yes}     & 1.00 & 1.18 & 1.14\\
NH13R10E20T100S   & 1D & 10 & 20MeV &  0.92 & 100 & {\bf Yes}     & 1.49 & 1.18 & 1.12\\
\hline                             
NH13R20E13T100S   & 1D & 20 & 13MeV &  0.66 & 100 & No            & ---  & 1.18 & --- \\
NH13R20E13T150S   & 1D & 20 & 13MeV &  0.66 & 150 & No            & ---  & 1.21 & --- \\
NH13R20E13T200S   & 1D & 20 & 13MeV &  0.66 & 200 & No            & ---  & 1.25 & --- \\
NH13R20E14T100S   & 1D & 20 & 14MeV &  0.88 & 100 & No            & ---  & 1.18 & --- \\
NH13R20E14T150S   & 1D & 20 & 14MeV &  0.88 & 150 & No            & ---  & 1.21 & --- \\
NH13R20E14T200S   & 1D & 20 & 14MeV &  0.88 & 200 & No            & ---  & 1.25 & --- \\
NH13R20E15T100S   & 1D & 20 & 15MeV &  1.16 & 100 & {\bf Yes}     & 0.97 & 1.18 & 1.15\\
NH13R20E15T150S   & 1D & 20 & 15MeV &  1.16 & 150 & {\bf Yes}     & 0.54 & 1.21 & $<1.24$ \\
NH13R20E15T200S   & 1D & 20 & 15MeV &  1.16 & 200 & {\bf Yes}     & 0.47 & 1.25 & $<1.26$ \\
NH13R20E21T100S   & 1D & 20 & 21MeV &  4.47 & 100 & {\bf Yes}     & 5.56 & 1.18 & 1.07\\
NH13R20E22T100S   & 1D & 20 & 22MeV &  5.39 & 100 & {\bf Yes}     & 6.50 & 1.18 & 1.07\\
\hline                             
NH13R28E13T100S   & 1D & 28 & 13MeV &  1.29 & 100 & No            & ---  & 1.18 & --- \\
NH13R29E13T100S   & 1D & 29 & 13MeV &  1.38 & 100 & No            & ---  & 1.18 & --- \\
NH13R30E11T100S   & 1D & 30 & 11MeV &  0.76 & 100 & No            & ---  & 1.18 & --- \\
NH13R30E11T150S   & 1D & 30 & 11MeV &  0.76 & 150 & No            & ---  & 1.21 & --- \\
NH13R30E11T200S   & 1D & 30 & 11MeV &  0.76 & 200 & No            & ---  & 1.25 & --- \\
NH13R30E12T100S   & 1D & 30 & 12MeV &  1.07 & 100 & No            & ---  & 1.18 & --- \\
NH13R30E12T150S   & 1D & 30 & 12MeV &  1.07 & 150 & No            & ---  & 1.21 & --- \\
NH13R30E12T200S   & 1D & 30 & 12MeV &  1.07 & 200 & No            & ---  & 1.25 & --- \\
NH13R30E13T100S   & 1D & 30 & 13MeV &  1.48 & 100 & {\bf Yes}     & 0.85 & 1.18 & $<1.19$ \\
NH13R30E13T150S   & 1D & 30 & 13MeV &  1.48 & 150 & No            & ---  & 1.21 & --- \\
NH13R30E13T200S   & 1D & 30 & 13MeV &  1.48 & 200 & No            & ---  & 1.25 & --- \\
NH13R30E14T100S   & 1D & 30 & 14MeV &  1.99 & 100 & {\bf Yes}     & 1.58 & 1.18 & 1.12\\
NH13R30E14T150S   & 1D & 30 & 14MeV &  1.99 & 150 & {\bf Yes}     & 0.98 & 1.21 & 1.19\\
NH13R30E14T200S   & 1D & 30 & 14MeV &  1.99 & 200 & {\bf Yes}     & 0.68 & 1.25 & 1.22\\
NH13R30E15T100S   & 1D & 30 & 15MeV &  2.62 & 100 & {\bf Yes}     & 2.27 & 1.18 & 1.10\\
NH13R30E15T150S   & 1D & 30 & 15MeV &  2.62 & 150 & {\bf Yes}     & 1.43 & 1.21 & 1.16\\
NH13R30E15T200S   & 1D & 30 & 15MeV &  2.62 & 200 & {\bf Yes}     & 0.93 & 1.25 & 1.22\\
NH13R30E20T100S   & 1D & 30 & 20MeV &  8.28 & 100 & {\bf Yes}     & 6.84 & 1.18 & 1.07\\
\hline                             
NH13R40E11T100S   & 1D & 40 & 11MeV &  1.35 & 100 & No            & ---  & 1.18 & --- \\
NH13R50E11T100S   & 1D & 50 & 11MeV &  2.10 & 100 & {\bf Yes}     & 0.86 & 1.18 & $<1.18$ \\
NH13R60E11T100S   & 1D & 60 & 11MeV &  3.03 & 100 & {\bf Yes}     & 1.48 & 1.18 & 1.12
\enddata
\label{tab:models}
\end{deluxetable*}

\subsubsection{The progenitor dependence}

In addition to the 13 $M_{\odot}$ progenitor by \cite{nomo88}, we are
going to investigate the progenitor dependence in 1D simulations. The
computed models are NH15 ($15M_\odot$) \citep{nomo88}, s15s7b2
($15M_\odot$) \citep{woos95}, s15.0 ($15M_\odot$), s20.0
($20M_\odot$), and s25.0 ($25M_\odot$) \citep{woos02}, which are
listed in Table \ref{tab:progenitor}.  The first sets of characters
for these models indicate the progenitors as,
\begin{itemize}
\item NH: \citep{nomo88}
\item WW: \citep{woos95}
\item WHW: \citep{woos02}
\end{itemize}

\begin{figure}[tbp]
\includegraphics[width=0.5\textwidth]{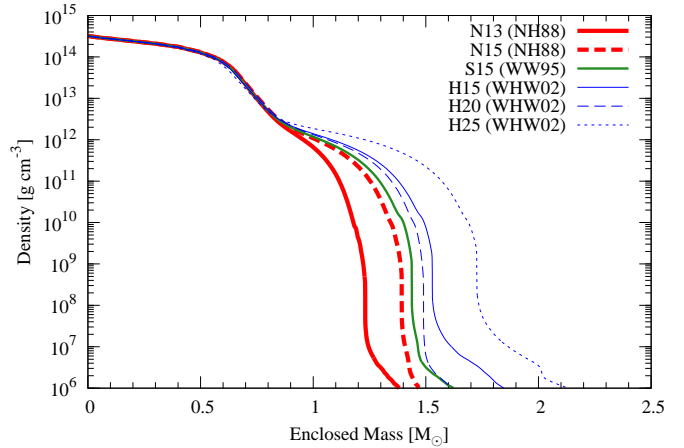}
\caption{Density profiles of investigated progenitors 100 ms after the
  bounce as functions of the enclosed mass. }
\label{fig:progenitor_den}
\end{figure}

Figure \ref{fig:progenitor_den} depicts density profiles of these
progenitors 100 ms after the bounce as a function of the enclosed mass
($M_r$). It can be seen that the density profiles for $M_r\lesssim 0.8
M_\odot$ are almost insensitive to the progenitor masses despite the
difference in the pre-collapse phase \citep[see, e.g., Figure 1
  of][]{burr07a}. On the other hand, the profiles of $M_r\gtrsim
0.8M_\odot$ differ between progenitors so that the critical heating
rates and $E_\mathrm{diag}^\infty$ are expected to be different also.
In Figure \ref{fig:progenitor_den}, the envelope of WHW25 is shown to
be thickest, while the envelope of NH13 is thinnest.

\begin{figure}[tbp]
\includegraphics[width=0.5\textwidth]{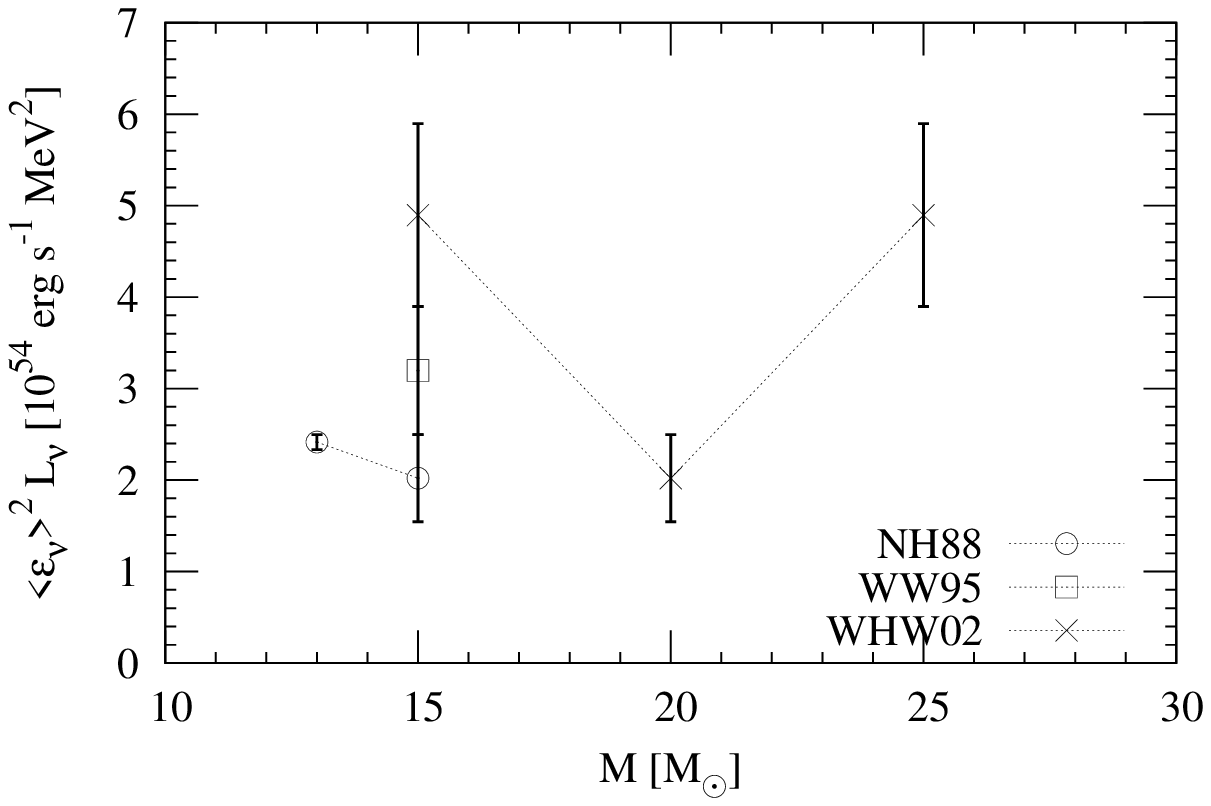}
\caption{The critical heating rate,
  $\bracket{\epsilon_{\nu_x}}^2L_{\nu_x}$, as a function of the
  progenitor mass, $M$. Circles are progenitors from \cite{nomo88},
  the square is from \cite{woos95}, and crosses are from
  \cite{woos02}, respectively.  The error bars represent the distance
  between the last failing and the first exploding model in our grid
  of models. The symbols locate at the centers of error bars. The
  error bar is small for model NH13 because we calculated a more
  refined grid of models for $13M_\odot$ progenitor (Table
  \ref{tab:models}) than for the 15-25$M_\odot$ progenitors (Table
  \ref{tab:progenitor}).  }
\label{fig:progenitor}
\end{figure}

Figure \ref{fig:progenitor} shows the critical heating rates as a
function of the progenitor masses. In agreement with intuition, the
critical heating rate for models WHW25 and NH13 belongs to the high
and low ends, respectively.  However, the critical heating rate for
model WHW20 is almost the same as the one for model NH13 although the
envelope of model WHW20 is much thicker than model NH13 (see Figure
\ref{fig:progenitor_den}). Our results show that the critical heating
rate is indeed affected by the envelope mass, however, the relation is
not one-to-one.  It is also interesting to note that the critical
heating rates for $15M_\odot$ progenitors of WW15, WHW15 and NH15, are
different by a factor of $\sim 3$, which may send us a clear message
that the accurate knowledge of supernova progenitors is also pivotal
to pin down the supernova mechanism.

\begin{figure}[tbp]
\includegraphics[width=0.5\textwidth]{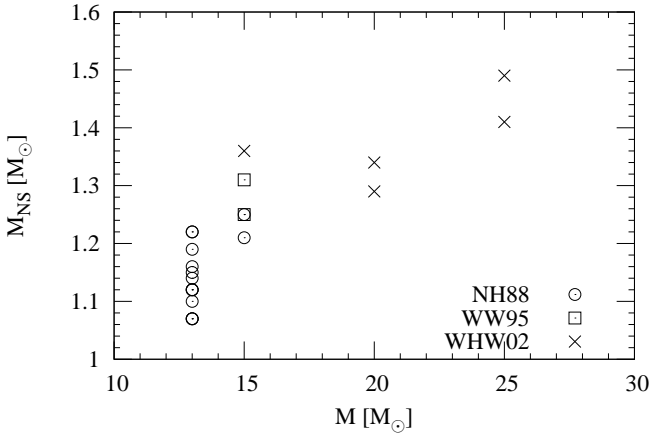}
\caption{The final NS masses as a function of the progenitor mass,
  $M$. Circles are progenitors from \cite{nomo88}, squares are from
  \cite{woos95}, and crosses are from \cite{woos02}, respectively.}
\label{fig:progenitor_ns}
\end{figure}

The integrated masses with $\rho\ge 10^{10}$ g cm$^{-3}$ for $t=t_s$
and $t=\infty$ are listed in the last two lines in Table
\ref{tab:progenitor} and Figure \ref{fig:progenitor_ns}.  The
tendencies are the same as found with NH13.  As for model WHW25, we
obtain results with $E_\mathrm{diag}^\infty>10^{51}$ erg and
$M_{10}^\infty>1.4 M_\odot$, simultaneously.

\begin{deluxetable*}{lccccccccc}
\tablecaption{Progenitor dependence}
\tablecolumns{10}
\tablewidth{0pc}
\tabletypesize{\scriptsize}
\tablehead{
\colhead{Model} & \colhead{Dimension} & \colhead{$R_\nu$} & \colhead{$\bracket{\epsilon_{\nu_x}}$} & \colhead{$L_\nu$} & \colhead{$t_s$}  & \colhead{Explosion} & \colhead{$E_\mathrm{diag}^ \infty$} & \colhead{$M_{10}^{t=t_s}$} & \colhead{$M_{10}^\infty$} \\
\colhead{} & \colhead{} & \colhead{[km]} & \colhead{[MeV]}  & \colhead{[10$^{52} $erg s$^{-1}$]} & \colhead{[ms]} & \colhead{} & \colhead{[$10^{51}$ erg]} & \colhead{[$M_\odot$]} & \colhead{[$M_\odot$]}
}
\startdata
NH15R30E11T100S   & 1D & 30 & 11MeV &  0.76 & 100 & No            & ---  & 1.34 & --- \\
NH15R30E12T100S   & 1D & 30 & 12MeV &  1.07 & 100 & No            & ---  & 1.34 & --- \\
NH15R30E13T100S   & 1D & 30 & 13MeV &  1.48 & 100 & {\bf Yes}     & 0.65 & 1.34 & $<1.38$ \\
NH15R30E14T100S   & 1D & 30 & 14MeV &  1.99 & 100 & {\bf Yes}     & 2.17 & 1.34 & 1.25\\
NH15R30E15T100S   & 1D & 30 & 15MeV &  2.62 & 100 & {\bf Yes}     & 3.73 & 1.34 & 1.21\\
\hline                             
WW15R30E11T100S   & 1D & 30 & 11MeV &  0.76 & 100 & No            & ---  & 1.40 & --- \\
WW15R30E12T100S   & 1D & 30 & 12MeV &  1.07 & 100 & No            & ---  & 1.40 & --- \\
WW15R30E13T100S   & 1D & 30 & 13MeV &  1.48 & 100 & No            & ---  & 1.40 & --- \\
WW15R30E14T100S   & 1D & 30 & 14MeV &  1.99 & 100 & {\bf Yes}     & 1.94 & 1.40 & 1.31\\
WW15R30E15T100S   & 1D & 30 & 15MeV &  2.62 & 100 & {\bf Yes}     & 3.41 & 1.40 & 1.25\\
\hline                             
WHW15R30E11T100S  & 1D & 30 & 11MeV &  0.76 & 100 & No            &  --- & 1.49 & --- \\
WHW15R30E12T100S  & 1D & 30 & 12MeV &  1.07 & 100 & No            &  --- & 1.49 & --- \\
WHW15R30E13T100S  & 1D & 30 & 13MeV &  1.48 & 100 & No            & ---  & 1.49 & --- \\
WHW15R30E14T100S  & 1D & 30 & 14MeV &  1.99 & 100 & No            & ---  & 1.49 & --- \\
WHW15R30E15T100S  & 1D & 30 & 15MeV &  2.62 & 100 & {\bf Yes}     & 3.55 & 1.49 & 1.36\\
\hline                             
WHW20R30E11T100S  & 1D & 30 & 11MeV &  0.76 & 100 & No            & ---  & 1.45 & --- \\
WHW20R30E12T100S  & 1D & 30 & 12MeV &  1.07 & 100 & No            & ---  & 1.45 & --- \\
WHW20R30E13T100S  & 1D & 30 & 13MeV &  1.48 & 100 & {\bf Yes}     & 0.99 & 1.45 & --- \\
WHW20R30E14T100S  & 1D & 30 & 14MeV &  1.99 & 100 & {\bf Yes}     & 2.20 & 1.45 & 1.34\\
WHW20R30E15T100S  & 1D & 30 & 15MeV &  2.62 & 100 & {\bf Yes}     & 3.61 & 1.45 & 1.29\\
\hline                             
WHW25R30E12T100S  & 1D & 30 & 12MeV &  1.07 & 100 & No            & ---  & 1.69 & --- \\
WHW25R30E13T100S  & 1D & 30 & 13MeV &  1.48 & 100 & No            & ---  & 1.69 & --- \\
WHW25R30E14T100S  & 1D & 30 & 14MeV &  1.99 & 100 & No            & ---  & 1.69 & --- \\
WHW25R30E15T100S  & 1D & 30 & 15MeV &  2.62 & 100 & {\bf Yes}     & 0.73 & 1.69 & $<2.00$ \\
WHW25R30E16T100S  & 1D & 30 & 16MeV &  3.39 & 100 & {\bf Yes}     & 5.92 & 1.69 & 1.49\\
WHW25R30E17T100S  & 1D & 30 & 17MeV &  4.32 & 100 & {\bf Yes}     & 9.21 & 1.69 & 1.41
\enddata
\label{tab:progenitor}
\end{deluxetable*}

\subsection{Two-dimensional models}
Here we discuss the effects of spectral swapping in 2D (axisymmetric)
simulations.  Since our 2D simulations, albeit utilizing the IDSA
scheme, are still computationally expensive, it is not practicable to
perform a systematic survey in 2D as we have done in 1D
simulations. Looking at Figure \ref{fig:progenitor} again, we choose
models WHW15 \citep{woos02} and NH13 \citep{nomo88}, whose critical
heating rate belong to the high and low ends, respectively.

\subsubsection{2D without spectral swapping}
The basic hydrodynamic picture is the same with 1D before the
shock-stall (e.g., till $\lesssim 10$ ms after bounce). After that,
convection as well as SASI sets in between the stalled shock and the
gain radius, which leads to the neutrino-heated shock revival for
model NH13 (e.g., \citet{suwa10}).  While for model WHW15, the
position of the stalled shock, following several oscillations, begins
to shrink at $\gtrsim 400$ ms after bounce.

Even after the shock revival, it should be emphasized that the shock
propagation for model NH13 is the so-called ``passive'' one
\citep{bura06}.  This means that the amount of the mass ejection is
smaller than the accretion in the post-shock region of the expanding
shock (see motions of mass shells in the post-shock region of Figure 1
in \citet{suwa10}).  Some regions have a positive local energy
(Eq. (\ref{eq:Ediag})), but the volume integrated value is quite as
small as $\lesssim 10^{50}$ erg at the maximum.  In order to reverse
the passive shock into an active one it is most important to energize
the explosion in some way.  Using these two progenitors that produce a
very weak explosion (model NH13) and do not show even a shock revival
(model WHW15), we hope to explore how the dynamics would change when
the spectral swapping is switched on.

\subsubsection{2D with spectral swapping}  
Table \ref{tab:2D} shows a summary for our 2D models, in which the
last character of each model (A) indicates ``Axisymmetry''. Models
NH13A and WHW15A are 2D models without spectral swapping for NH13 and
WHW15, respectively.

As in 1D, the onset of the spectral swapping is taken to be $t_s=100$
ms after bounce.  At this time, model NH13 shows the onset of the
gradual shock expansion with a small diagnostic energy of
$E_\mathrm{diag}\sim 3\times 10^{49}$ erg, and the shock radius is
located at $\sim 300$ km.  As for model WHW15, there is no region with
a positive local energy (e.g., Eq. (\ref{eq:Ediag})) and the shock
radius is $\sim 200$ km. The density profile for this model is
essentially same as the one in the 1D counterpart (see Figure
\ref{fig:progenitor_den}) but with small angular density modulations
due to convection.

In Figure \ref{fig:Ediag}, red filled circles represent
$E_\mathrm{diag}^\infty$ for model NH13. It can be seen that the
critical heating rate to obtain $E_\mathrm{diag}^\infty \sim 10^{51}$
erg is smaller for 2D than the corresponding 1D counterparts (compare
the heating rates for $\bracket{\epsilon_{\nu_x}}^2 L_{\nu_x} \sim
2.2\times10^{54}~\mathrm{MeV^2~erg~s^{-1}}$). In fact, models with
$\bracket{\epsilon_{\nu_x}}^2 L_{\nu_x}\lesssim
2.2\times10^{54}~\mathrm{MeV^2~erg~s^{-1}}$ fail to explode in 1D, but
succeed in 2D (albeit with a relatively small $E_\mathrm{diag}^\infty$
less than $10^{51}$ erg). As opposed to 1D, it is rather difficult in
2D to determine a critical heating rate due to the stochastic nature
of the explosion triggered by SASI and convection. In our limited set
of 2D models, the critical heating rate is expected to be close to
$\bracket{\epsilon_{\nu_x}}^2 L_{\nu_x}\sim
1.5\times10^{54}~\mathrm{MeV^2~erg~s^{-1}}$, below which the shock
does not revive (e.g., $\bracket{\epsilon_{\nu_x}}^2 L_{\nu_x}
\lesssim 1\times10^{54}~\mathrm{MeV^2~erg~s^{-1}}$, is the lowest end
in the horizontal axis in the figure).

As seen from Figure \ref{fig:Ediag}, $E_\mathrm{diag}^\infty$ becomes
visibly larger for 2D than 1D especially for a smaller
$\bracket{\epsilon_{\nu_x}}^2 L_{\nu_x}$.  As the heating rates become
larger, the difference between 1D and 2D becomes smaller because the
shock revival occurs almost in a spherically symmetric way (before
SASI and convection develop non-linearly).  In Table 3, it is
interesting to note that model NH13R30E11T100A fails to explode, while
we observed the shock-revival for the corresponding model without the
spectral swapping (model NH13A).  This is because the heating rate of
model NH13R30E11T100A is smaller than NH13A due to the small
$\bracket{\epsilon_{\nu_x}}$, which can make it more difficult to
trigger $\nu_x$ explosions. On the other hand, if the energy gain due
to the swap is high enough (i.e., for models with greater than E12 in
Table \ref{tab:2D}), the swap can facilitate explosions.

\begin{figure*}[tbp]
\includegraphics[width=\textwidth]{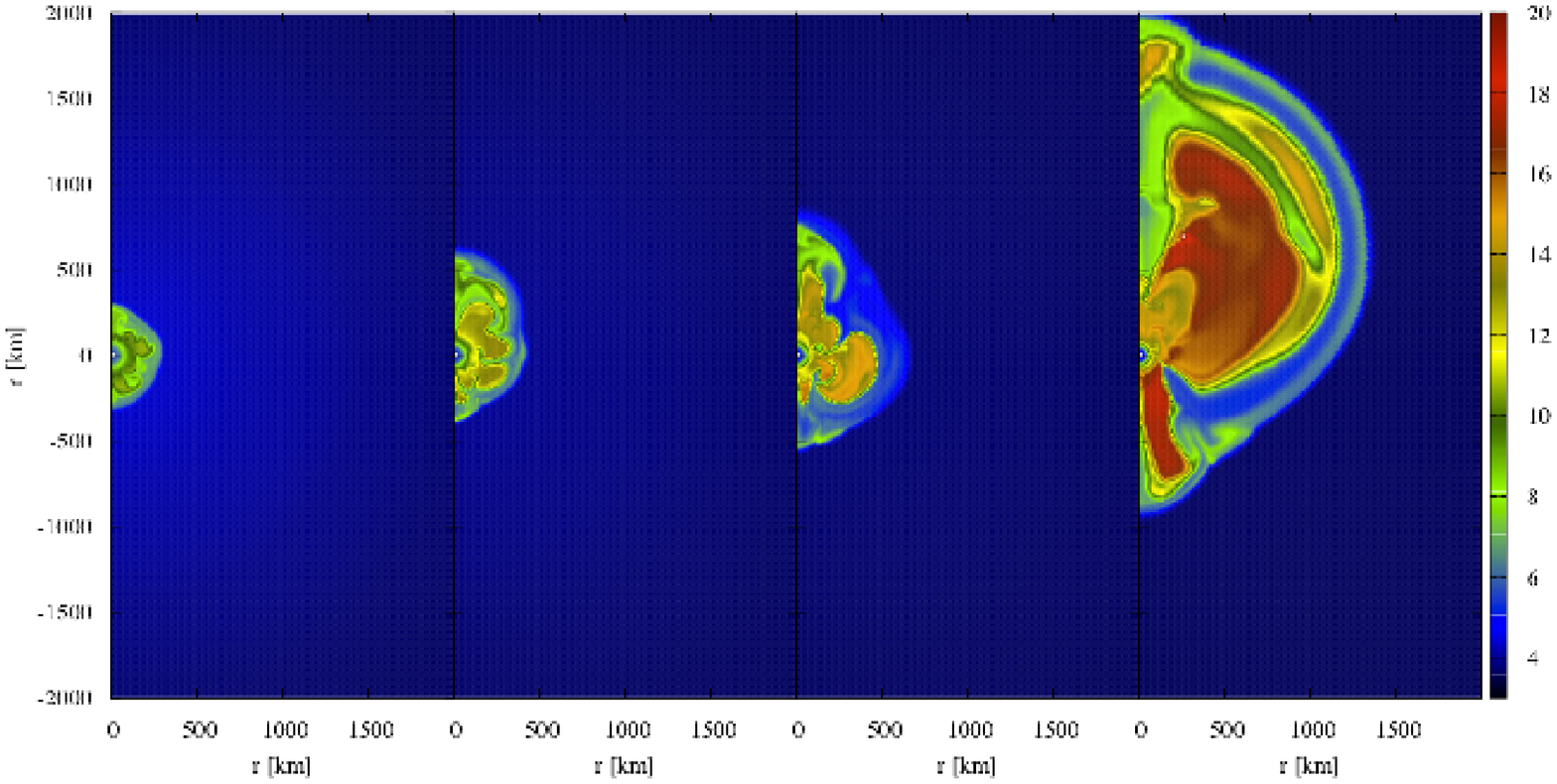}
\includegraphics[width=\textwidth]{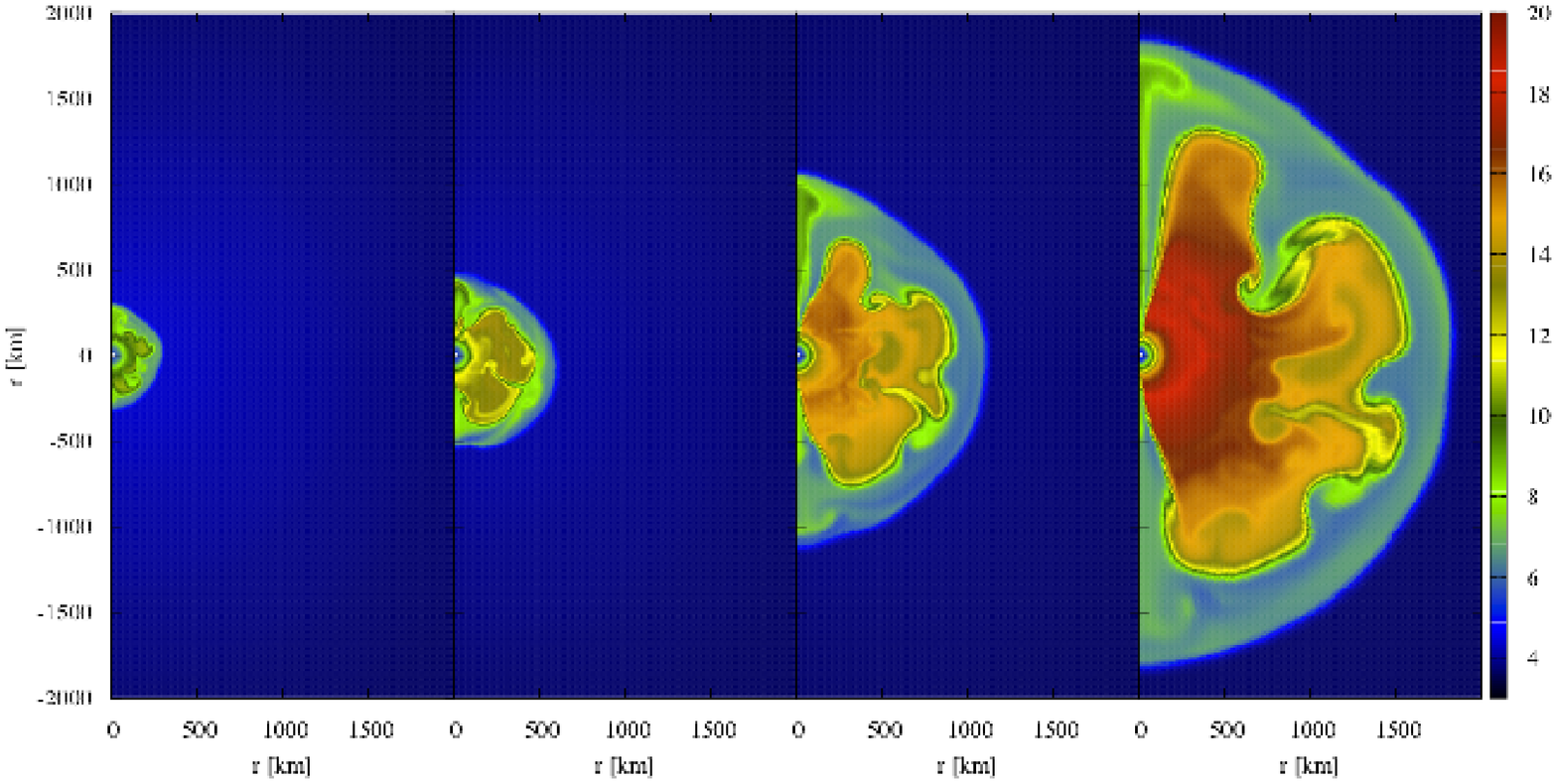}
\caption{Time evolution of the entropy distributions. {\it Top}: NH13A
  without the spectral swapping for 100, 200, 300, and 450 ms after
  bounce from left to right.  {\it Bottom}: NH13R30E13T100A for 100,
  150, 200, 250 ms after bounce (corresponding to 0, 50, 100, 150 ms
  after the onset of the spectral swapping.)}
\label{fig:2d_time_ev}
\end{figure*}

Figure \ref{fig:2d_time_ev} depicts the entropy distributions for
models NH13A (top panel) and NH13R30E13T100A (bottom panel). It can be
seen that model NH13A shows a unipolar-like explosion \citep[see
  also][]{suwa10}, while model NH13R30E13A explodes rather in a
spherical manner as mentioned above. Model NH13A experiences several
oscillations aided by SASI and convection before explosion, while the
stalled shock for model NH13R30E13T100A, turns into expansion shortly
after the onset of the spectral swapping.  In fact, the shapes of hot
bubbles behind the expanding shock are shown to be barely changing
with time (bottom panel), which indicates a quasi-homologous expansion
of material behind the revived shock.

\begin{figure}[tbp]
\includegraphics[width=0.5\textwidth]{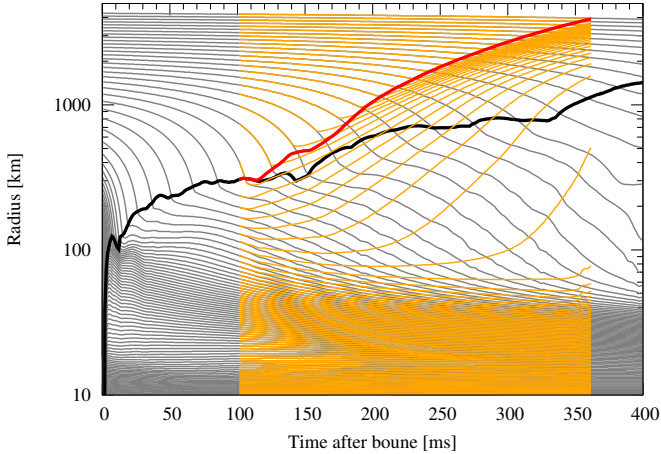}
\caption{Time evolution of mass shells for NH13A (thin-gray lines) and
  NH13R30E13T100A (thin-orange lines). Black and red thick lines
  represent the shock position at the north pole.}
\label{fig:mass_shell_2d}
\end{figure}

Figure \ref{fig:mass_shell_2d} shows the time evolution of mass shells
for models NH13A (thin-gray lines) and NH13R30E13T100A (thin-orange
lines). Black and red thick lines represent the shock position at the
north pole for each models.  The mass shells for model NH13A continue
to accrete to the PNS, since the shock {\it passively} expands
outwards as already mentioned.  Due to this continuing mass accretion,
the remnant for this model would be a black hole instead of a neutron
star.  On the other hand, model NH13R30E13T100A shows a mass ejection
with a definite outgoing momentum in the postshock region so that the
remnant could be a neutron star. Unfortunately however, we cannot
predict the final outcome due to the limited simulation time.  A
long-term simulation recently done in 1D (e.g., \citet{fisc10}) should
be indispensable also for our 2D case. This is, however, beyond the
scope of this paper.

Here let us discuss a validity of the parameters for the spectral swap
that we have assumed so far. For example, the criteria of explosion
for model NH13R30E12T100A was $L_{\nu_x}\approx 1.07\times10^{51}$erg
s$^{-1}$ and $\bracket{\epsilon_{\nu_x}}\approx12$ MeV. These values
are even smaller than the typical values obtained in 1D Boltzmann
simulations (e.g., \citet{lieb04}), which show $L_{\nu_x}\approx
2\times 10^{52}$ erg s$^{-1}$ and
$\sqrt{\bracket{\epsilon_{\nu_x}^2}}\approx 20$ MeV
(i.e. $\bracket{\epsilon_{\nu_x}}\approx 14$ MeV with a vanishing
chemical potential) earlier in the postbounce phase.  Therefore the
spectral swapping, if it would work as we have assumed, may be a
potential to assist explosions.

It should be noted that the critical heating rate in this study might
be too small due to the approximation of the light-bulb scheme. In
this scheme, we can include the geometrical effect of the finite size
of the neturinosphere as in Eq. (\ref{eq:f}), but can not include the
back reaction by the matter, i.e. the absorption of neutrino. Some
fraction of neutrinos, in fact, are absorbed in the gain region and
the neutrino luminosity decreases with the radius. We omit this effect
in this study so that the heating rate might be overestimated in the
simulation with the spectral swapping. Thus, the fully consistent
simulation including the spectral swapping is necessary for more
realistic critical heating rate, which is beyond the scope of this
study.

\begin{figure}[tbp]
\includegraphics[width=.45\textwidth]{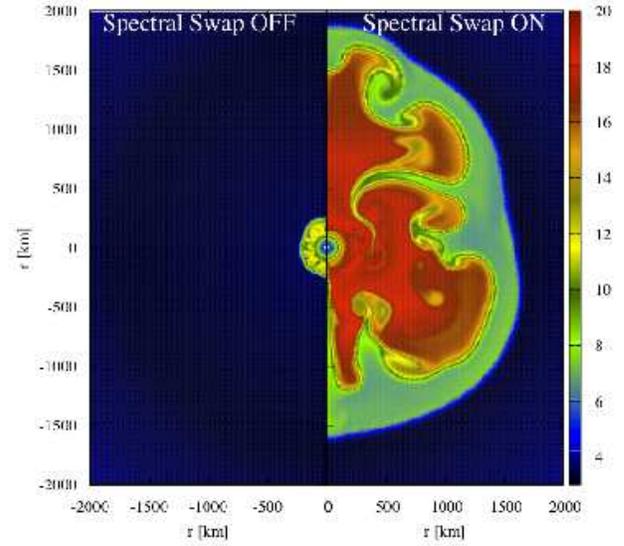}
\caption{The entropy distributions of WHW15A (left) and
  WHW15R30E15T100A (right) for 220 ms after the bounce.}
\label{fig:whw15}
\end{figure}

Finally we discuss the 15 $M_{\odot}$ progenitor labeled by WHW15.  As
mentioned, this progenitor fails to explode without spectral swapping
even in 2D\footnote{This is consistent with a very recent result by
  \citet{ober11}, who performed 2D simulations of model WHW15 with
  spectral neutrino transport}. Figure \ref{fig:whw15} shows the
entropy distributions of WHW15A (left; nonexploding) and
WHW15R30E15T100A (right; exploding) for 220 ms after the bounce
(corresponding to 120 ms after $t_s$ for model WHW15R30E15T100A). The
model with $R_{\nu_x}=30$ km and $\bracket{\epsilon_{\nu_x}}=14$ MeV
does not explode in 1D but explodes in 2D (compare Table
\ref{tab:progenitor} and \ref{tab:2D}). Again, the
mulitidimensionality helps the onset of explosion.  The critical
heating rate in 2D is in the range of
$2.5\le\bracket{\epsilon_{\nu_x}}^2L_{\nu_x}/(10^{54}$ MeV$^2$ erg
s$^{-1})\le 3.9$, while it is $3.9\le\bracket{\epsilon_{\nu_x}}^2
L_{\nu_x}/(10^{54}$ MeV$^2$ erg s$^{-1})\le 5.9$ in 1D.  Therefore the
critical heating rate in 2D can be by a factor $\sim 2$ smaller than
in 1D. In 2D, a critical $\nu_x$ luminosity and average energy to
obtain explosion are $L_{\nu_x}\sim 2\times 10^{52}$ erg s$^{-1}$ and
$\bracket{\epsilon_{\nu_x}}\sim 14$ MeV (corresponding to
$\sqrt{\bracket{\epsilon_{\nu_x}^2}}\sim 20$ MeV), which are close to
the results obtained in a 1D Boltzmann simulation \citep{sumi05} for a
15 $M_{\odot}$ progenitor\footnote{Note that the progenitor employed
  in \citet{sumi05} is WW95, so that the direct comparison may not be
  fair.  However, the critical heating rate in 1D for WW15 is smaller
  than WHW15 (Figure \ref{fig:progenitor}) and the mass of the
  envelope is thicker for WHW15 than WW15 (Figure
  \ref{fig:progenitor_den}). This indicates that our discussion above
  seems to be quite valid, although we really need 1D results for
  WHW15 to draw a more solid conclusion.}. The diagnostic energy as
well as the estimated remnant masses are listed in the last three
columns in Table \ref{tab:2D}.  $E_\mathrm{diag}^\infty$ (as well as
$M_\mathrm{diag}^\infty$) for exploding models is shown to be larger
than the model series of NH13.  As a result, some of the 2D models for
WHW15 produce strong explosions ($E_\mathrm{diag}^\infty \sim 10^{51}$
erg), while simultaneously leaving behind a remnant of 1.34--1.52
$M_\sun$. We think that it is only a solution accidentally found by
our parametric explosion models. However again, the critical heating
rates that require to assist the neutrino-driven explosion via the
spectral swapping are never far away from the ones obtained in the
Boltzmann simulations. We hope that our exploratory results may give a
momentum to supernova theorists to elucidate the effects of collective
neutrino oscillations in a more consistent manner.

\begin{deluxetable*}{lccccccccc}
\tablecaption{2D simulations}
\tablecolumns{10}
\tablewidth{0pc}
\tabletypesize{\scriptsize}
\tablehead{
\colhead{Model} & \colhead{Dimension} & \colhead{$R_\nu$} & \colhead{$\bracket{\epsilon_{\nu_x}}$} & \colhead{$L_\nu$} & \colhead{$t_s$}  & \colhead{Explosion} & \colhead{$E_\mathrm{diag}^ \infty$} & \colhead{$M_{10}^{t=t_s}$} & \colhead{$M_{10}^\infty$} \\
\colhead{} & \colhead{} & \colhead{[km]} & \colhead{[MeV]}  & \colhead{[10$^{52} $erg s$^{-1}$]} & \colhead{[ms]} & \colhead{} & \colhead{[$10^{51}$ erg]} & \colhead{[$M_\odot$]} & \colhead{[$M_\odot$]}
}
\startdata
NH13A & 2D & --- & --- & --- & --- & {\bf Yes}     & $\sim 0.1$ (oscillating) & --- & ---\\
NH13R30E11T100A   & 2D & 30 & 11MeV &  0.76 & 100 & No            &  --- & 1.18 & ---\\
NH13R30E12T100A   & 2D & 30 & 12MeV &  1.07 & 100 & {\bf Yes}     &  0.45 & 1.18 & $<1.23$\\
NH13R30E13T100A   & 2D & 30 & 13MeV &  1.48 & 100 & {\bf Yes}     &  1.03 & 1.18 & $<1.18$\\
NH13R30E15T100A   & 2D & 30 & 15MeV &  2.62 & 100 & {\bf Yes}     &  2.33 & 1.18 & 1.10\\
\hline
WHW15A & 2D & --- & --- & --- & --- & No & --- & --- & ---\\
WHW15R30E13T100A   & 2D & 30 & 13MeV &  1.48 & 100 & No            &  --- & --- & ---\\
WHW15R30E14T100A   & 2D & 30 & 14MeV &  1.99 & 100 & {\bf Yes}     & 1.96 & 1.48 & $<1.52$\\
WHW15R30E15T100A   & 2D & 30 & 15MeV &  2.62 & 100 & {\bf Yes}     & 3.79 & 1.48 & 1.34
\enddata
\label{tab:2D}
\end{deluxetable*}

\section{Summary and Discussion}

We performed a series of one- and two-dimensional hydrodynamic
simulations of core-collapse supernovae with spectral neutrino
transport via the IDSA scheme.  To model the spectral swapping which
is one of the possible outcomes of the collective neutrino
oscillations, we parametrized the onset time when the spectral swap
begins, the radius where the spectral swap takes place, and the
threshold energy above which the spectral interchange between
heavy-lepton neutrinos and electron/anti-electron neutrinos occurs. By
doing so, we systematically studied the shock evolution and the matter
ejection due to the neutrino heating enhanced by spectral swapping.
We also investigated the progenitor dependence using a suite of
progenitor models (13, 15, 20, and 25 $M_\odot$).  With these
computations, we found that there is a critical heating rate induced
by the spectral swapping to trigger explosions, which differs between
the progenitors.  The critical heating rate is generally smaller for
2D than 1D due to the multidimensionality that enhances the neutrino
heating efficiency \citep[see also][]{jank96}.  The remnant masses can
be determined by the mass ejection driven by the neutrino heating,
which range in 1.1-1.5$M_\odot$ depending on the progenitors.  For our
2D model of the $15M_\odot$ progenitor, we found a set of the
parameters that produces an explosion with a canonical supernova
energy close to $10^{51}$ erg and at the same time leaves behind a
remnant mass close to $\sim 1.4 M_\odot$.  Our results suggest that
collective neutrino oscillations have the potential to solve the
supernova problem if they occurs. These effects should be explored in
a more self-consistent manner in hydrodynamic simulations.

Here it should be noted that the simulations in this paper are only a
very first step towards more realistic supernova modeling.  For the
neutrino transfer, we omitted the cooling of heavy lepton neutrinos
and the inelastic neutrino scattering by electrons. These omissions
lead to an overestimation of the diagnostic energy and also they
should relax the criteria for explosion. The ray-by-ray approximation
may lead to an overestimation of the directional dependence of the
neutrino anisotropies. A full-angle transport will give us a more
correct answer \citep[see][]{ott08,bran11}. Moreover, due to the
coordinate symmetry axis, the SASI develops preferentially along the
axis; it could thus provide a more favorable condition for the
explosion. As several exploratory simulations have been done recently
\citep[e.g.,][]{iwak08,sche08,nord10}, 3D supernova models are indeed
necessary also to pin down the outcomes of the spectral swapping.

Finally we briefly discuss whether the oscillation parameters taken in
this paper are really valid in views of recent work whose focus is on
clarifying the still-veiled nature of collective neutrino
oscillations.  Following \cite{duan10}, there are at least two
conditions for the onset of collective neutrino oscillations in the
case of inverted neutrino mass hierarchy.

\begin{figure}[tbp]
\includegraphics[width=0.5\textwidth]{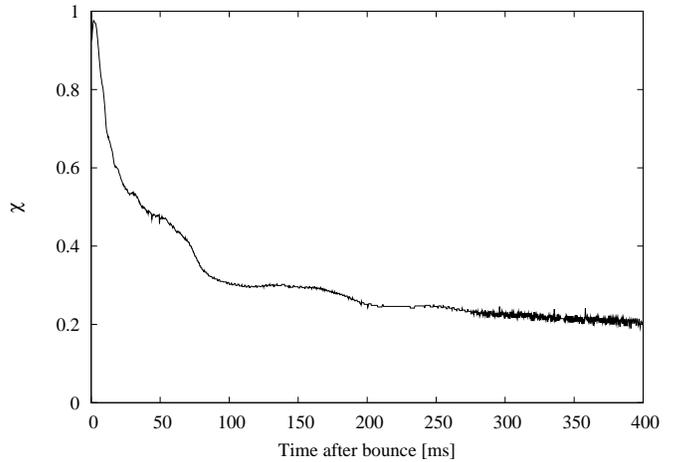}
\caption{ Time evolution of $\chi=F_{\nu_e}/F_{\bar\nu_e}-1$, where
  $F_{\nu_i}$ is the number flux of $\nu_i$.  }
\label{fig:chi}
\end{figure}

The first criteria should be satisfied in the so-called bipolar regime
of the collective oscillation. In the regime, the neutrino number
density should exceed the critical value,
\begin{eqnarray}
n_{\bar\nu_e, \rm crit}
&\simeq&\frac{1}{(\sqrt{1+\chi}-1)^2}\frac{\Delta m^2}{\sqrt{2}G_\mathrm{F}
\bracket{\epsilon_{\bar\nu_e}}}\nonumber\\
&\simeq& 1.4\times 10^{29} \mathrm{cm^{-3}}\left(\frac{0.2}{\chi}\right)^2\left(\frac{15~\mathrm{MeV}}{\bracket{\epsilon_{\bar\nu_e}}}\right),
\end{eqnarray}
where $\chi$ is the fractional excess of neutrinos over antineutrinos,
$\Delta m^2$ is the characteristic mass-squared splitting (a typical
value of $\sim 2.4\times 10^{-3}$ eV$^2$ is employed here), and
$G_\mathrm{F}$ is Fermi coupling constant. By using our simulation
results, we can estimate $\chi$ which is often treated as a parameter
(typically $\sim$0.01-0.25) so far.  The following estimation is given
in \cite{este07}, that is $\chi\simeq F_{\nu_e}/F_{\bar\nu_e}-1$ in
the case of vanishing $F_{\nu_x}$, where $F_{\nu_i}$ is the number
flux of $\nu_i$. From Figure \ref{fig:chi}, it can be seen that
$\chi\sim$ 0.2-0.3 for 100-400 ms after bounce.  Since the typical
number density in the post-shock region ($r\sim$200-300 km) can be
estimated as,
\begin{eqnarray}
n_{\bar\nu_e}&=&\frac{L_{\bar\nu_e}}{4\pi r^2 c\bracket{\epsilon_{\bar\nu_e}}}\nonumber\\
&\sim&1.1\times 10^{31}~\mathrm{cm}^{-3}
\left(\frac{L_{\bar\nu_e}}{10^{52}~\mathrm{erg~s^{-1}}}\right)
\left(\frac{100~\mathrm{km}}{r}\right)^2\nonumber\\
&\times&\left(\frac{15~\mathrm{MeV}}{\bracket{\epsilon_{\bar\nu_e}}}\right),
\end{eqnarray}
therefore, the first condition is satisfied\footnote{Even if $\chi$ is
  as small as $\chi \sim 0.01$ due to the inclusion of $\nu_x$, the
  criteria could be marginally satisfied.}.

The second criteria is related to the decoherence of collective
oscillations by matter. In order to overwhelm the suppression by the
decohenrence, the following condition should be satisfied
\begin{equation}
n_{\bar\nu_e,{\rm crit}} \sim n_e,
\end{equation}
where $n_e$ is the number density of electrons where the decoherence
takes place.  This is equivalent to,
\begin{equation}
Y_{\bar\nu_e,{\rm crit}} \sim Y_e.
\label{eq:Ynu}
\end{equation}
In our 1D simulation, $Y_{\bar\nu_e}\sim (0.1-0.2)\times Y_e$ for 100
km $\lesssim r\lesssim r_\mathrm{sh}$, where $r_\mathrm{sh}$ is the
shock radius\footnote{Outside the shock, $Y_{\bar\nu_e}>Y_e$ is
  achieved due to rapid density decrease.}. Since this condition is
barely satisfied, the collective oscillations in reality could modify
the spectrum to some extent between heavy-lepton neutrinos and
electron/anti-electron neutrinos, however the full swapping assumed in
this study may be exaggerated. Very recently\footnote{In fact they
  posted their papers on astro-ph after our submission.},
\cite{chak11a,chak11b} pointed out that the matter effect could fully
suppress the spectral swapping in the accretion phase using 1D
neutrino-radiation hydrodynamic simulation data of
\cite{fisc10}. However, the current understanding of the collective
oscillation is not completed and calculations in this field employ
several assumptions (e.g., single angle approximation) \citep[but see
  also][for more recent work]{dasg11}. To draw a robust conclusion,
one needs a more detailed study including the collective neutrino
flavor oscillation to the hydrodynamic simulations in a more
self-consistent manner, which we are going to challenge as a sequel of
this study.

\acknowledgements We thank to K. Sumiyoshi, H. Suzuki, S. Yamada,
T. Yoshida for stimulating discussions.  Numerical computations were
in part carried on XT4 at CfCA of the National Astronomical
Observatory of Japan. ML are supported by the Swiss National Science
Foundation under grant No. PP00P2-124879 and 200020-122287.  This
study was supported in part by the Japan Society for Promotion of
Science (JSPS) Research Fellowships (YS), the Grants-in-Aid for the
Scientific Research from the Ministry of Education, Science and
Culture of Japan (Nos. 19104006, 19540309 and 20740150), and HPCI
Strategic Program of Japanese MEXT.

\appendix
\section{Code Validity}
\subsection{Conservation of Energy}

\begin{figure}[htbp]
  \centering
  \includegraphics[width=.5\textwidth]{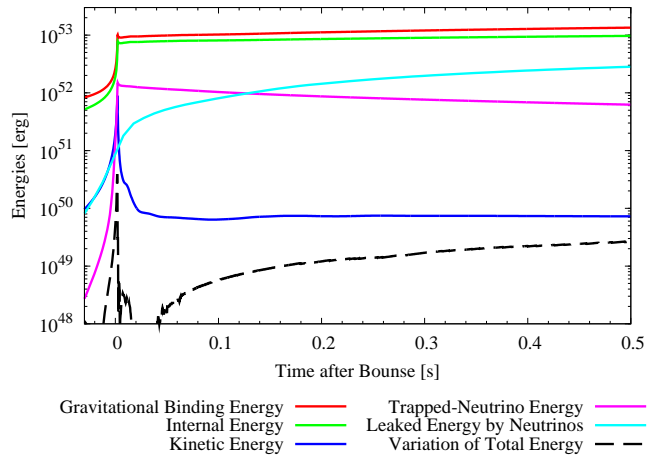}
  \caption{Time evolution of gravitational energy (red), internal
    energy (green line), kinetic energy (blue line), trapped-neutrino
    energy (magenta line), released energy by neutrinos (cyan line),
    and summation of these energies (black dashed line). These
    quantities are determined by integration with respect to volume
    included in our simulation except for released energy by neutrinos
    (magenta), which is $\int (L_{\nu_e}+L_{\bar\nu_e}) dt$. Since
    gravitational energy and total energy are negative, the absolute
    values are shown. The violation of total energy (dashed line)
    remains $< 3\times 10^{49}$ erg, which is $\sim 0.03\%$ of
    gravitational energy and internal energy after bounce ($\sim
    10^{53}$ erg).}
  \label{fig:ensy_cnsv}
\end{figure}

In this section, we demonstrate the conservation of physical
quantities using the spherical collapse model (NH13). Figure
\ref{fig:ensy_cnsv} depicts the evolution of total binding energy by
gravity (red line), total internal energy (green), total kinetic
energy (blue), total trapped-neutrino energy (magenta), total energy
leaked by neutrinos (cyan), and variation of overall energy (black
dashed), respectively. The gravitational energy and total energy are
negative and absolute values are shown. The gravitational energy and
internal energy dominate (with different sign) and reach $\sim
10^{53}$ erg soon after bounce. Despite such an enormous energy
change, the total energy varies only within $\sim 3\times10^{49}$ erg
so that the violation of energy conservation remains $< 0.03\%$.  The
energy of the trapped neutrinos decreases with the diffusion
timescale, which leads to the PNS cooling. The kinetic energy rapidly
drops because of the photodissociation of iron and the electron
capture ($\nu_e$ emission) that is consistent with the shock stall.
We have monitored these values in a 2D simulation and obtained a
similar level of energy conservation.

\clearpage
\subsection{Comparison with AGILE}

Here, we present the result of our numerical simulation in spherical
symmetry and compare with the result of AGILE-IDSA code
\citep{lieb09}. AGILE (Adaptive Grid with Implicit Leap Extrapolation)
is an implicit general relativistic hydrodynamics code that evolves
the Einstein equations based on conservative finite differencing on an
adaptive grid. We employ a one-dimensional version of our ZEUS-2D code
that has been developed to perform multidimensional supernova
simulations.

We compare the evolution of a $13 M_\odot$ star of \citet{nomo88} in
Newtonian gravity from precollapse model to 100 ms after bounce. We
find good agreement between the results of the ZEUS-2D and AGILE
during the early postbounce phase when the neutrino burst is launched
and the accretion shock expands to its maximum radius. The
hydrodynamic quantities are shown in following figures.

\begin{figure*}[htbp]
  \centering
  \includegraphics[width=.24\textwidth]{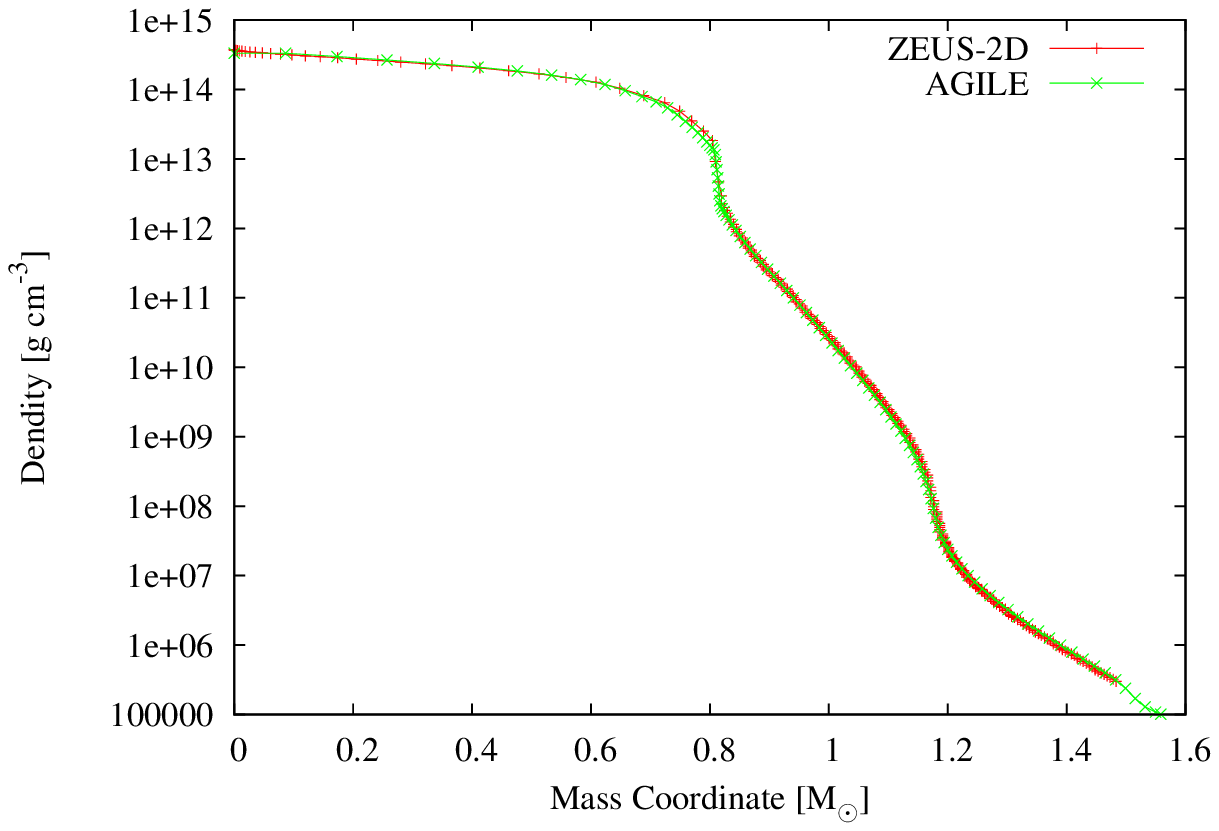}
  \includegraphics[width=.24\textwidth]{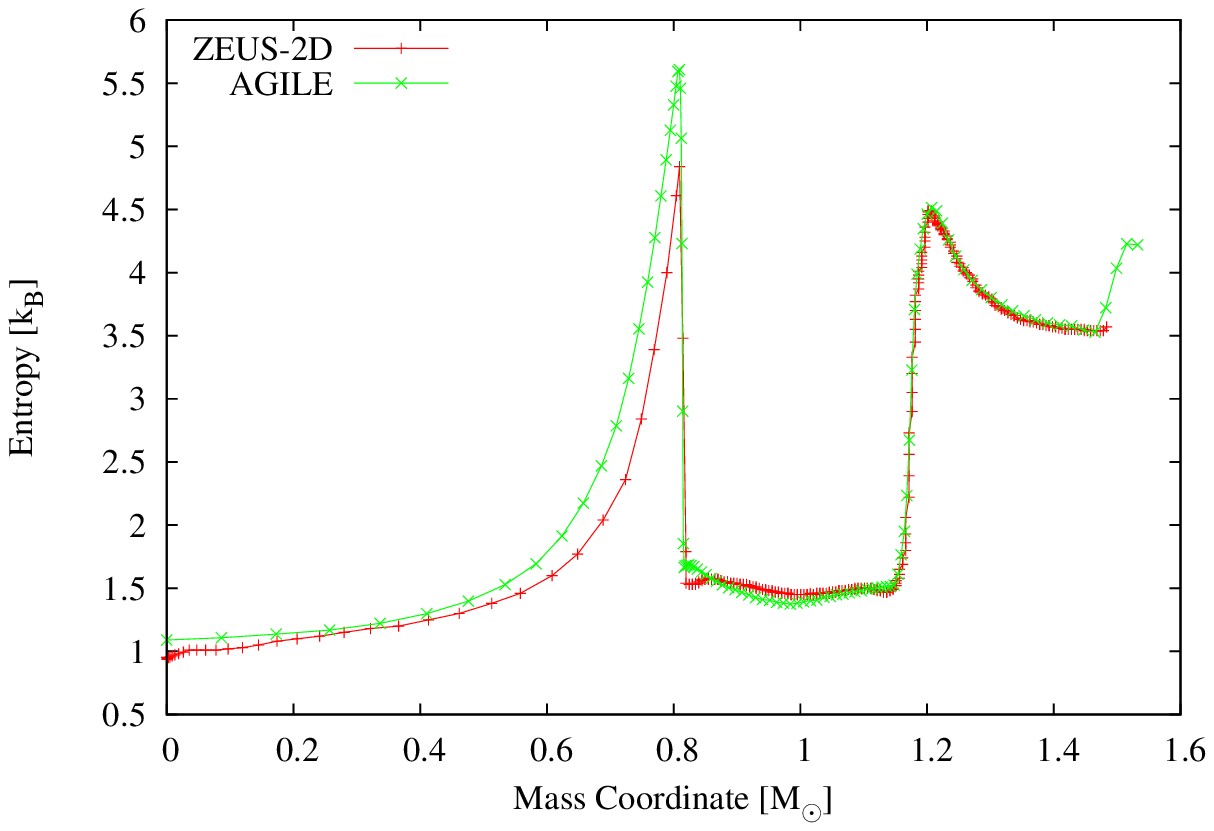}
  \includegraphics[width=.24\textwidth]{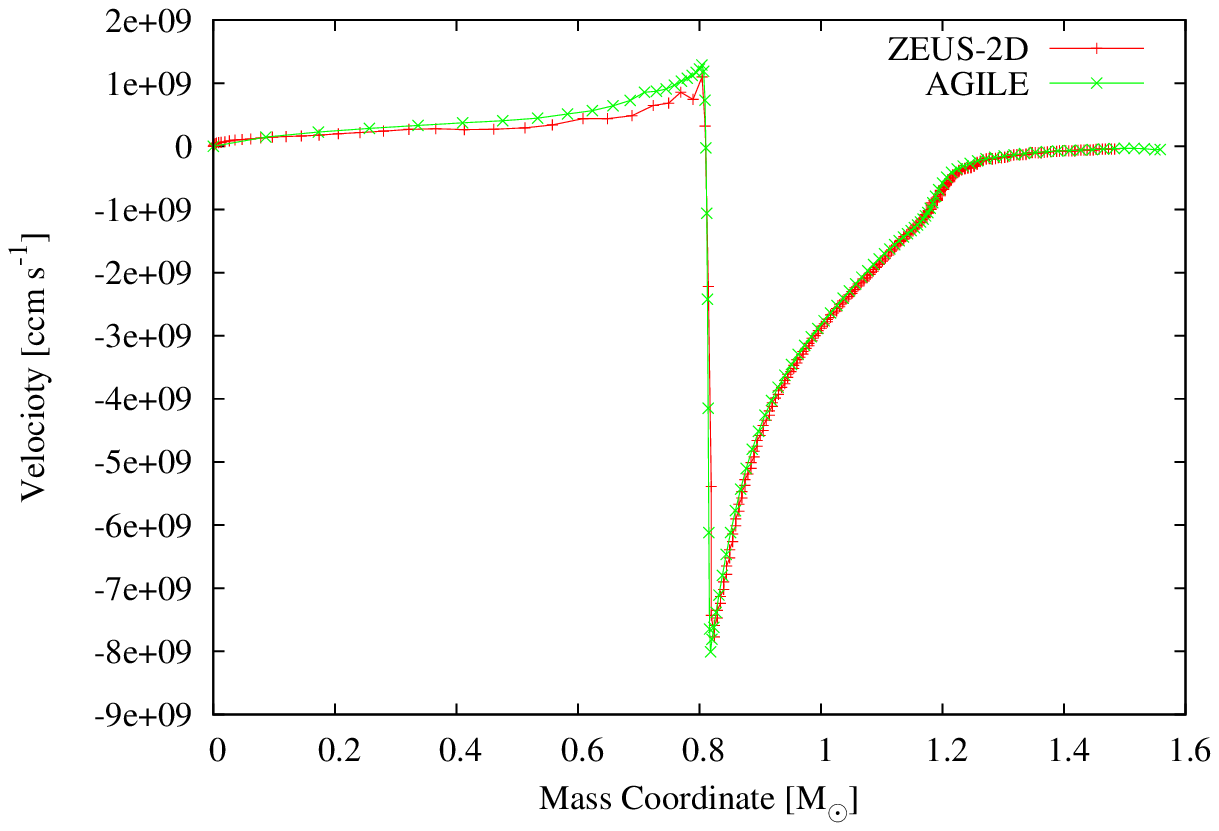}
  \includegraphics[width=.24\textwidth]{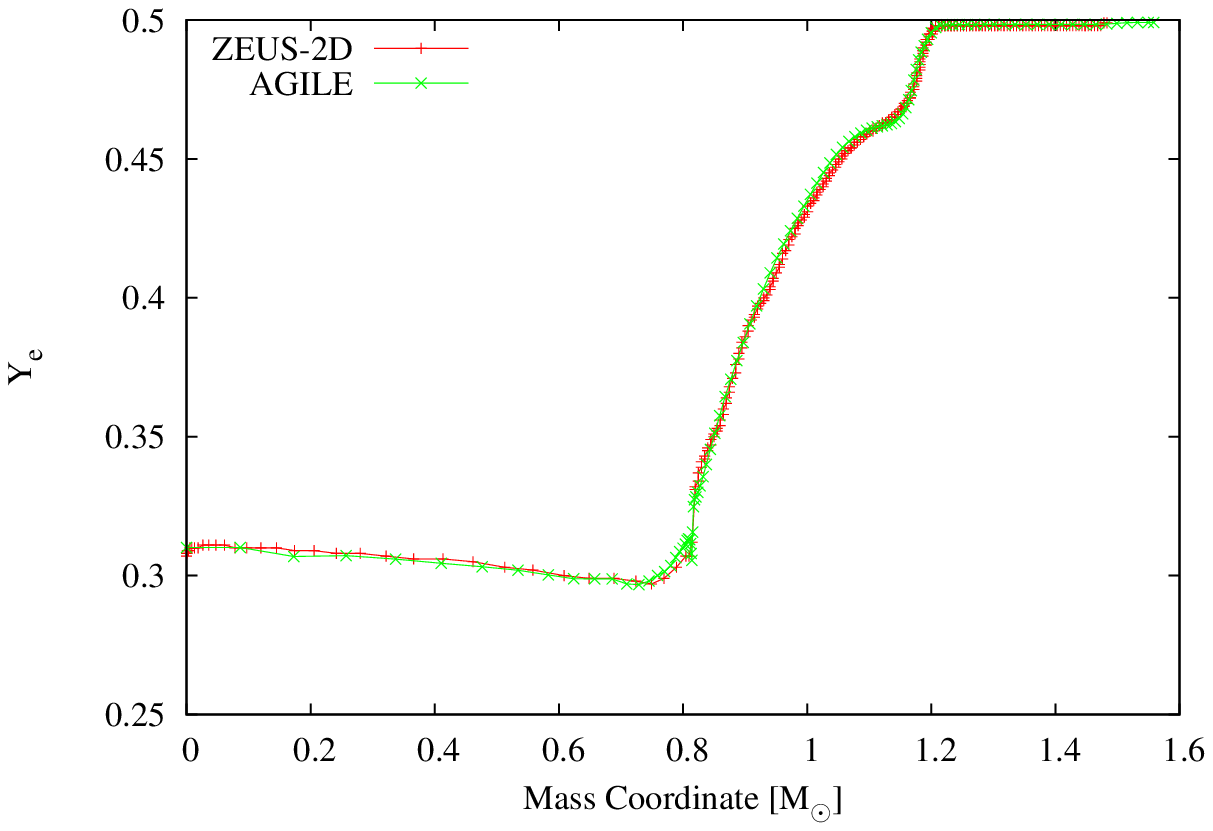}

  \caption{Density (left top), entropy (right top), velocity (left
    bottom), and electron fraction (right bottom) as a function of
    enclosed mass for the result of ZEUS-2D (red lines) and AGILE
    (green lines). The comparison is shown at the time just after the
    bounce. A difference is seen in the entropy profile, which comes
    from the difference of shock capturing scheme.}
  \label{fig:d1}
\end{figure*}

\begin{figure*}[htbp]
  \centering
  \includegraphics[width=.24\textwidth]{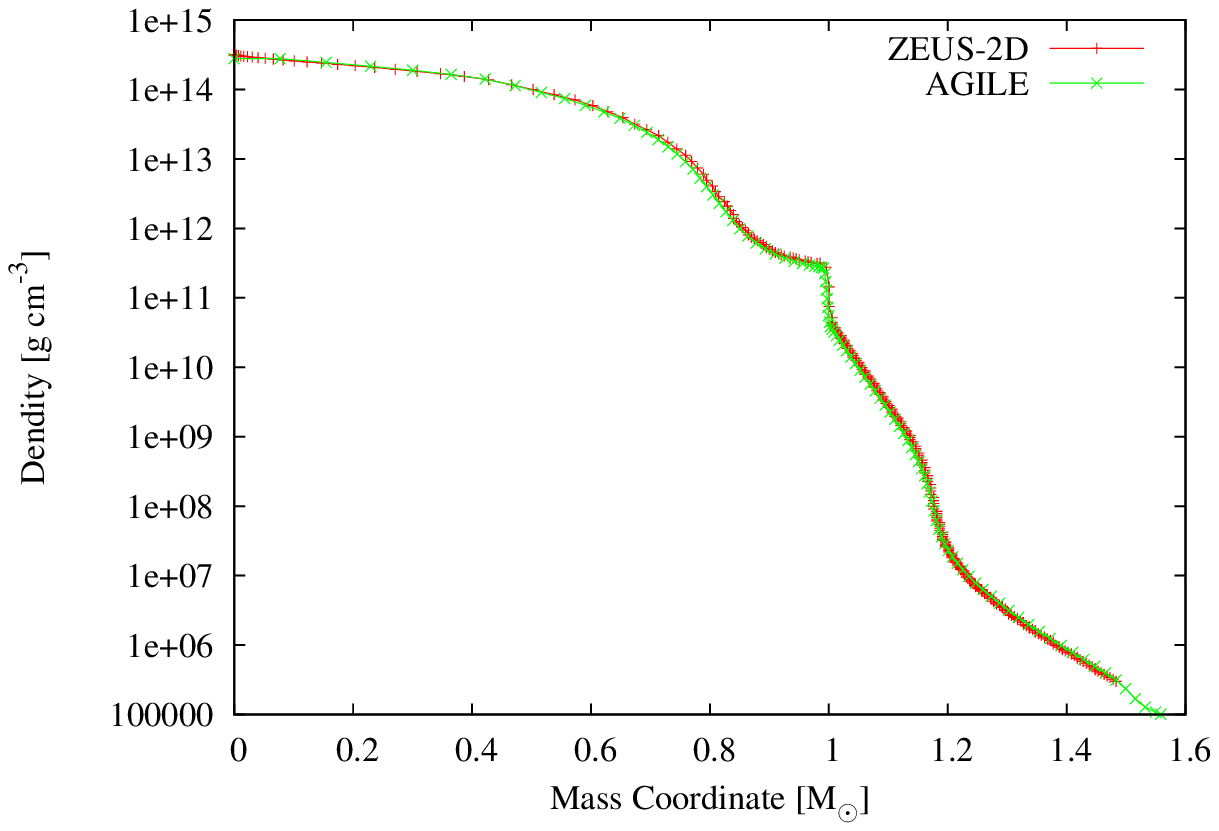}
  \includegraphics[width=.24\textwidth]{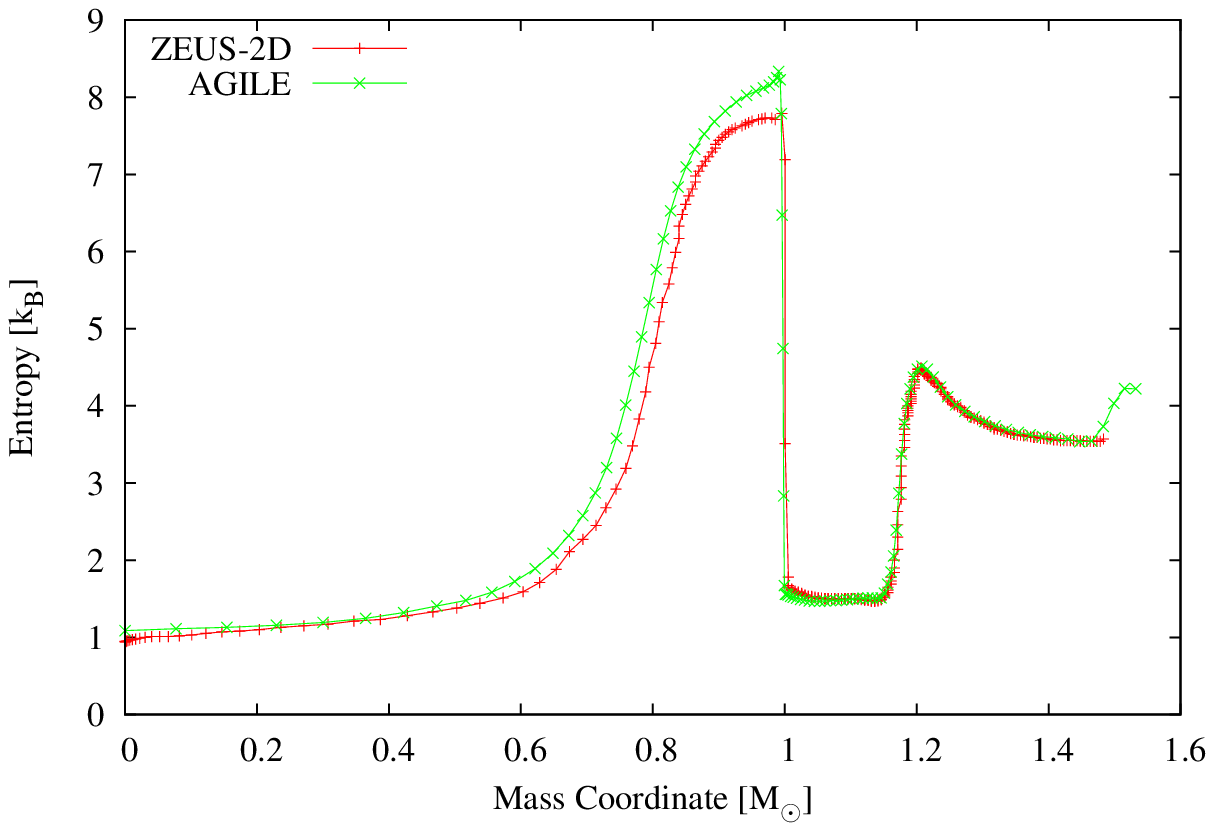}
  \includegraphics[width=.24\textwidth]{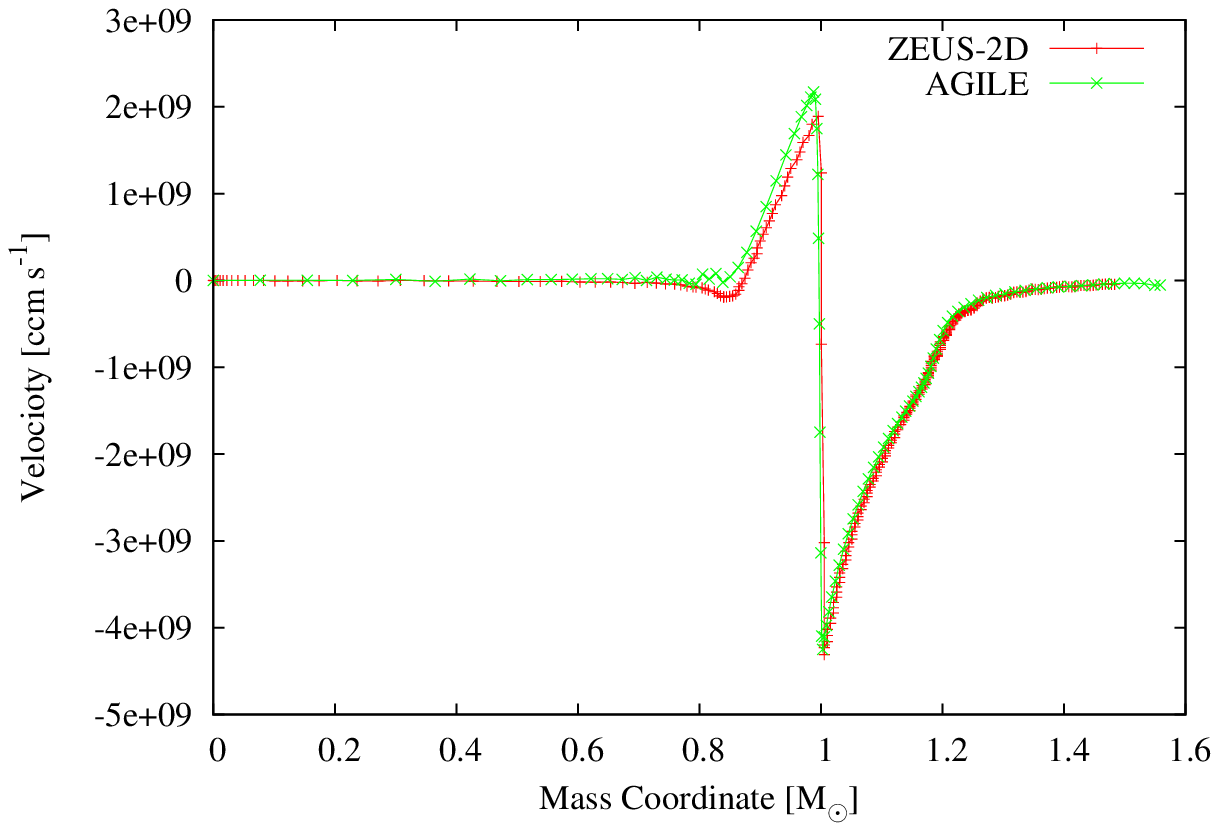}
  \includegraphics[width=.24\textwidth]{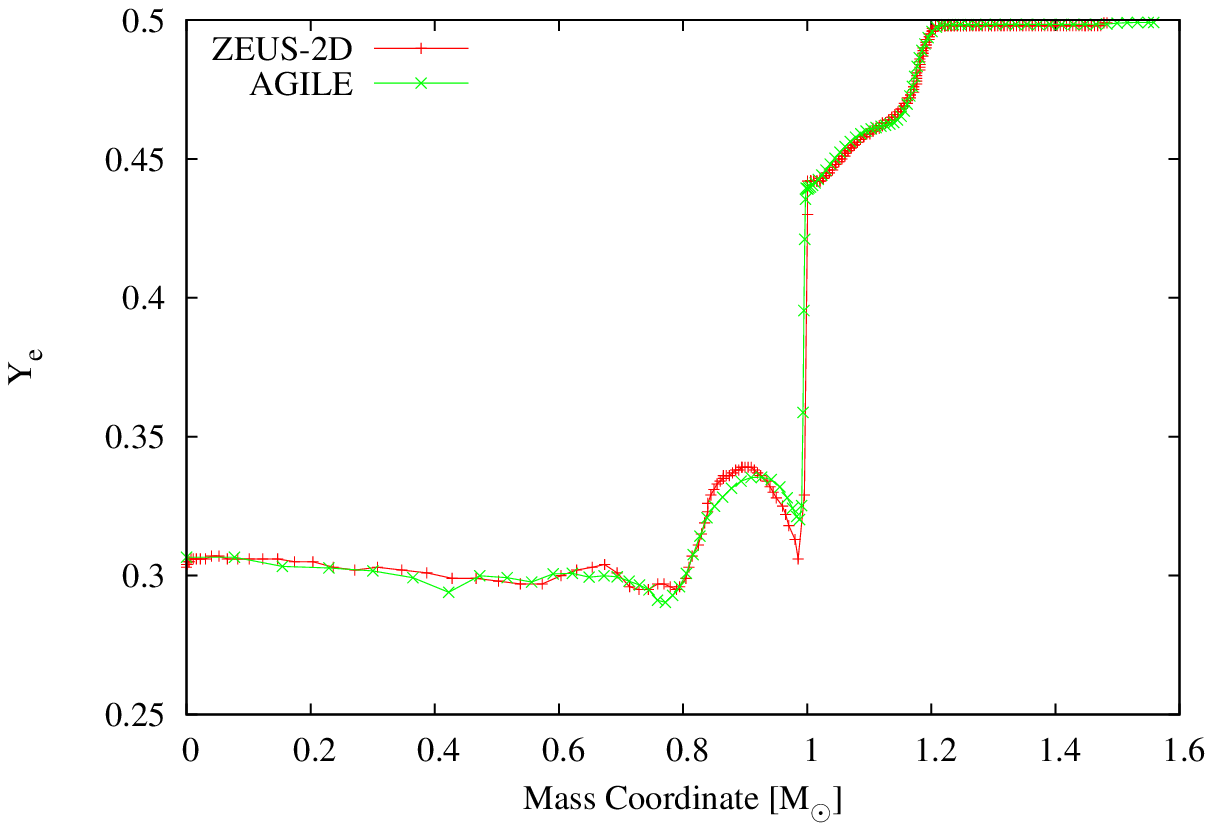}
  \caption{Same as Fig. \ref{fig:d1} but for the time at 1 ms after
    bounce.}
\end{figure*}

\begin{figure*}[htbp]
  \centering
  \includegraphics[width=.24\textwidth]{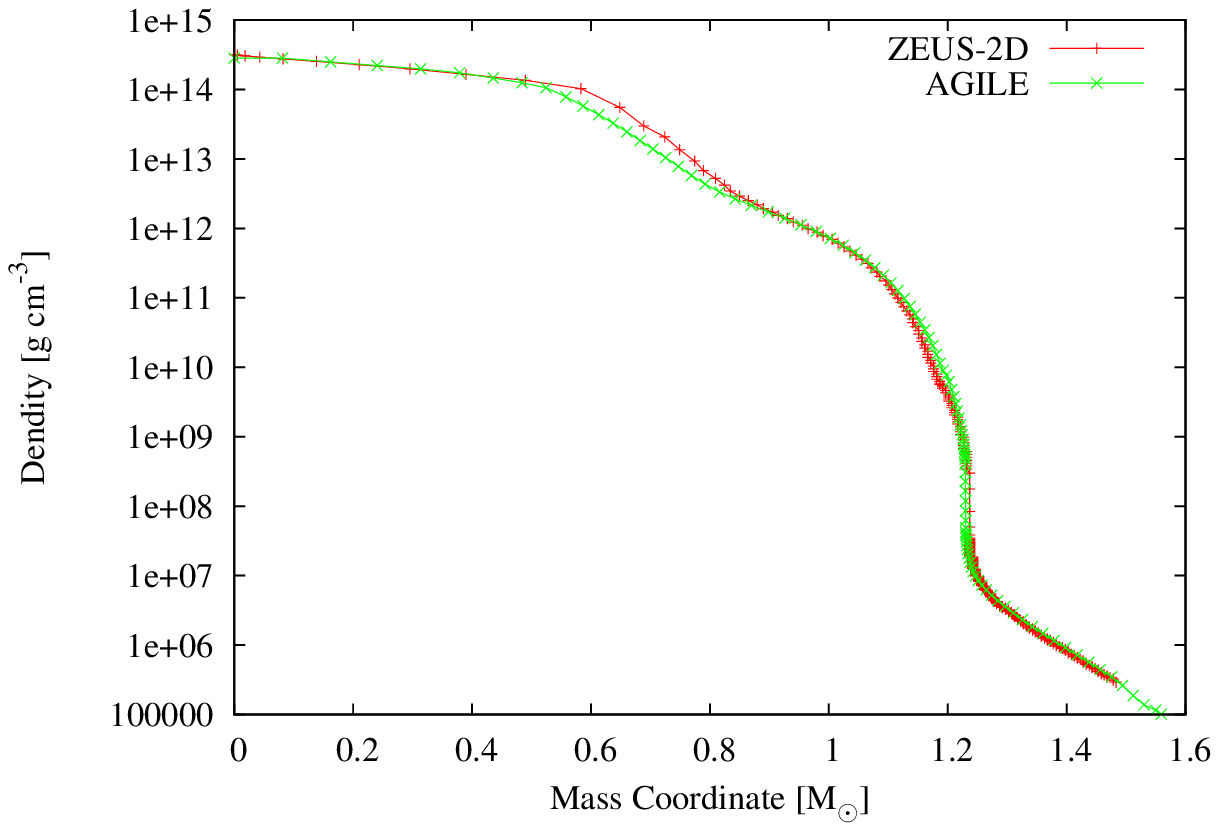}
  \includegraphics[width=.24\textwidth]{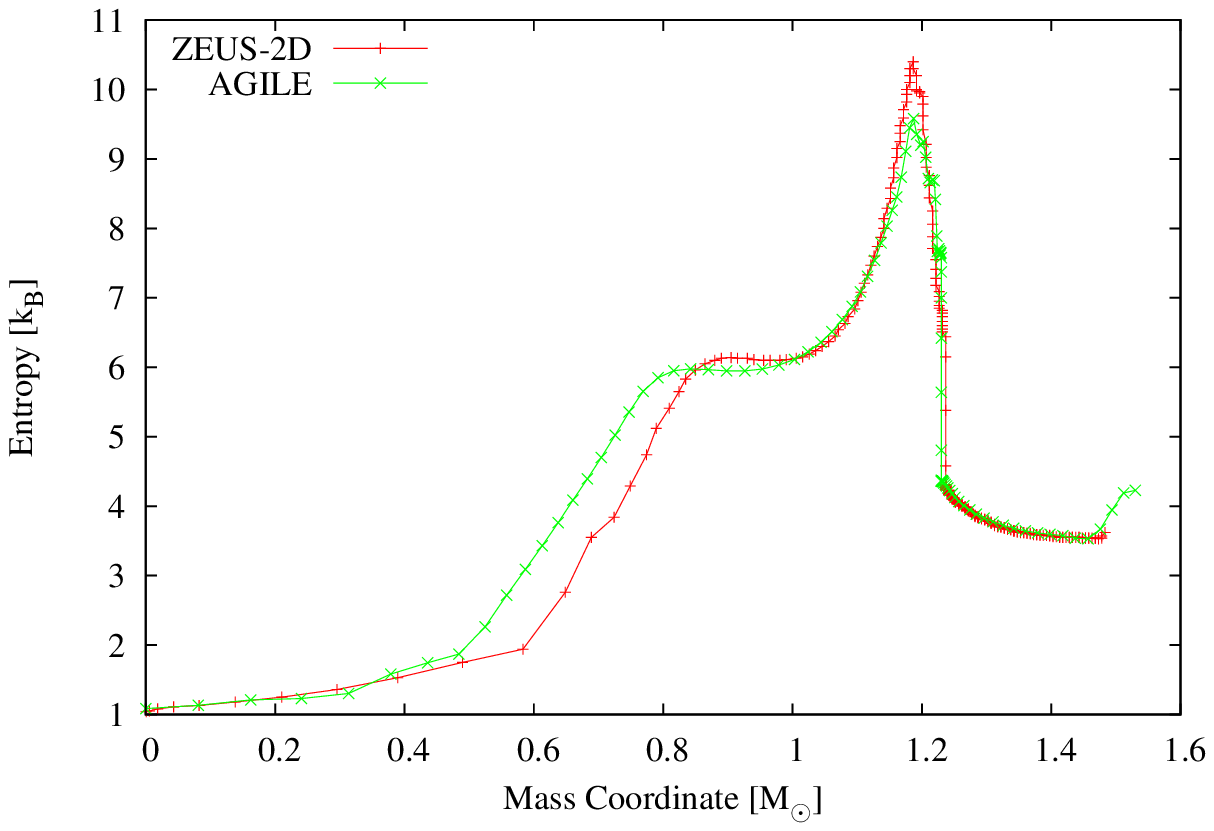}
  \includegraphics[width=.24\textwidth]{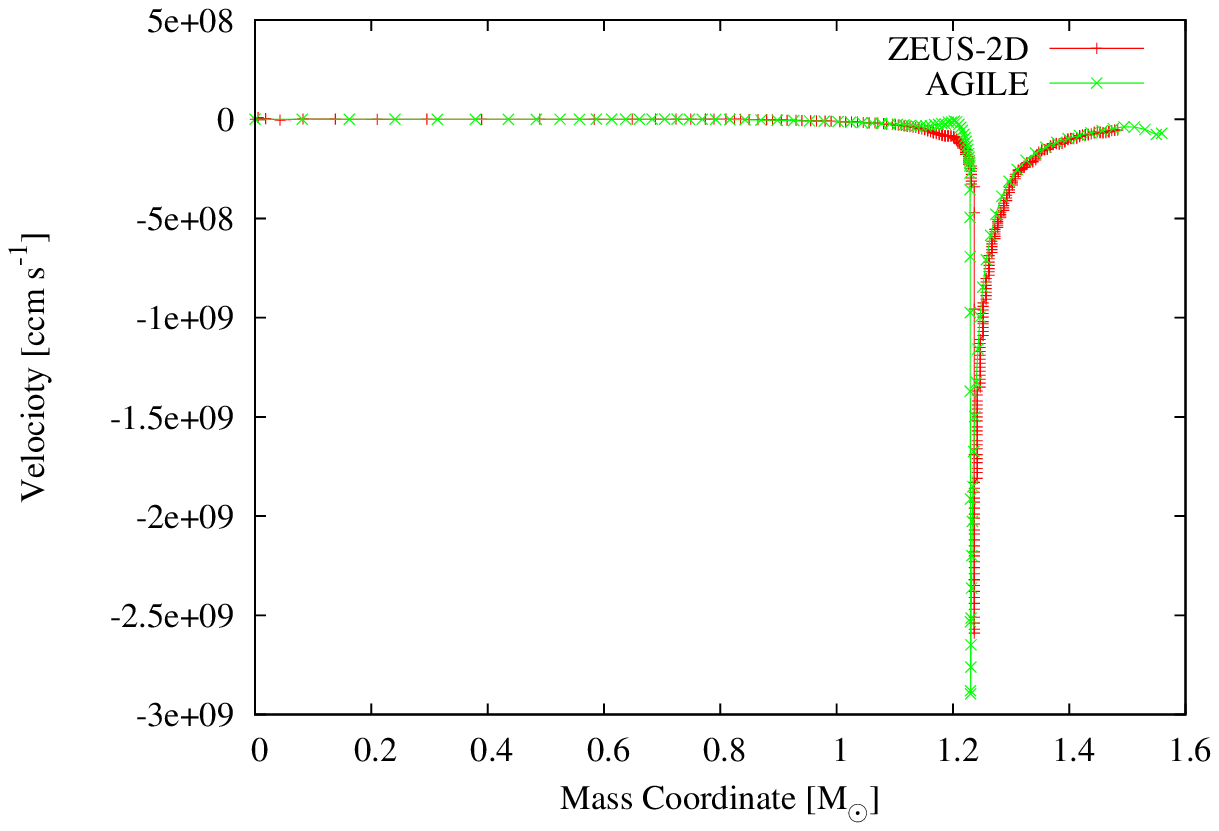}
  \includegraphics[width=.24\textwidth]{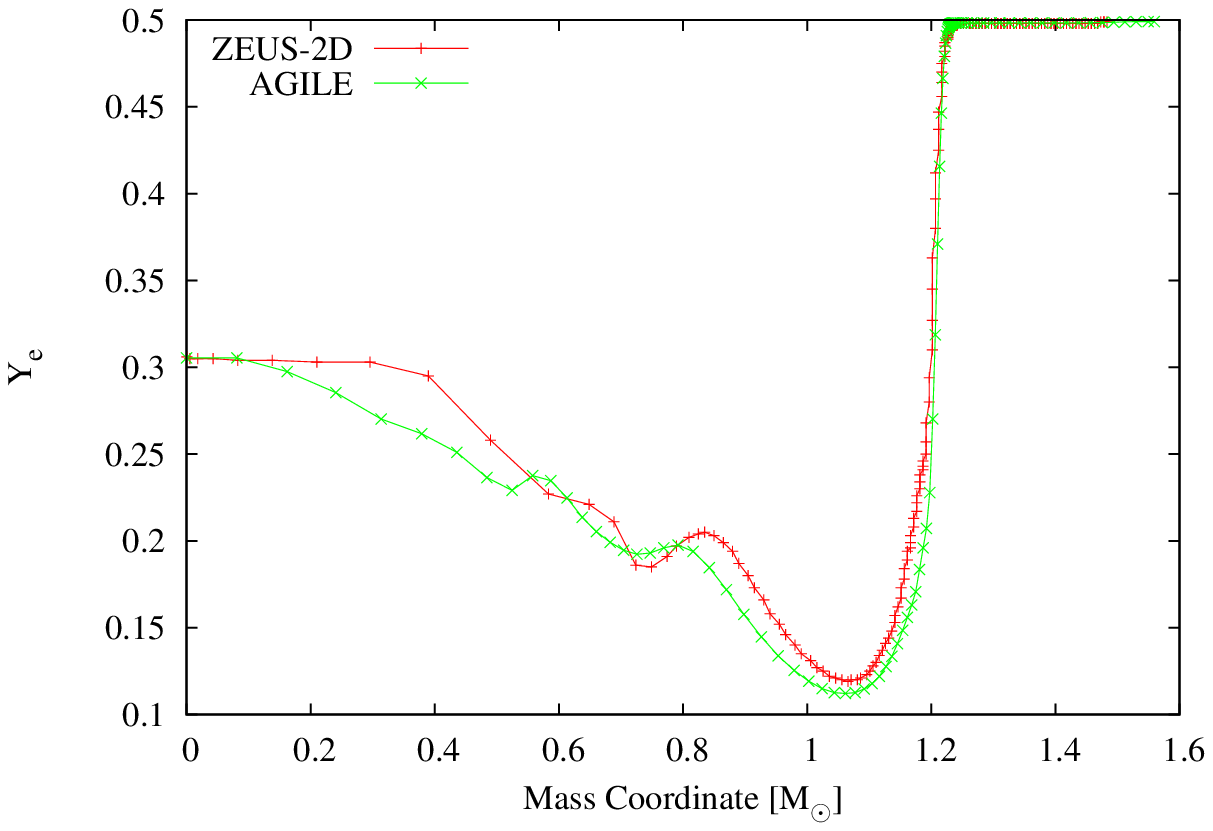}
  \caption{Same as Fig. \ref{fig:d1} but for the time at 100 ms after
    bounce.}
\end{figure*}

\end{document}